\documentclass[12pt,preprint]{aastex}

\begin{document}

\title{Clumpy Galaxies in GOODS and GEMS: Massive Analogs
of Local Dwarf Irregulars}

\author{Debra Meloy Elmegreen}
\affil{Vassar College, Dept. of Physics \& Astronomy, Box 745,
Poughkeepsie, NY 12604} \email{elmegreen@vassar.edu}

\author{Bruce G. Elmegreen}
\affil{IBM Research Division, T.J. Watson Research Center, P.O. Box
218, Yorktown Heights, NY 10598} \email{bge@watson.ibm.com}

\author{Max T. Marcus}
\affil{Vassar College, Dept. of Physics \& Astronomy, Box 745, Poughkeepsie, NY 12604}
\email{mamarcus@vassar.edu}

\author{Karlen Shahinyan}
\affil{Wesleyan University, Dept. of Astronomy, Middletown, CT 06459}
\email{kshahinyan@wesleyan.edu}

\author{Andrew Yau}
\affil{Vassar College, Dept. of Physics \& Astronomy, Box 745, Poughkeepsie, NY 12604}
\email{anyau@vassar.edu}

\author{Michael Petersen}
\affil{Colgate University, Dept. of Physics \& Astronomy, Hamilton, NY
13346} \email{mpetersen@students.colgate.edu}
\begin{abstract}
Clumpy galaxies in the GEMS and GOODS fields are examined for clues to
their evolution into modern spirals. The magnitudes of the clumps and
the surface brightnesses of the interclump regions are measured and
fitted to models of stellar age and mass.  There is an evolutionary
trend from clump clusters with no evident interclump emission to clump
clusters with faint red disks, to spiral galaxies of the flocculent or
grand design types.  Along this sequence, the interclump surface
density increases and the mass surface density contrast between the
clumps and the interclump regions decreases, suggesting a gradual
dispersal of clumps to form disks. Also along this sequence, the
bulge-to-clump mass ratios and age ratios increase, suggesting a
gradual formation of bulges.  All of these morphological types occur in
the same redshift range, indicating that the clump cluster morphology
is not the result of bandshifting. This redshift range also includes
clear examples of interacting galaxies with tidal tails and other
characteristic features, indicating that clump clusters, which do not
have these features, are not generally interacting. Comparisons to
local galaxies with the same rest wavelength and spatial resolution
show that clump clusters are unlike local flocculent and spiral
galaxies primarily because of the high clump/interclump contrasts in
the clump clusters. They bear a striking resemblance to local dwarf
Irregulars, however. This resemblance is consistent with a model in
which the clumpy morphology comes from gravitational instabilities in
gas with a high turbulent speed compared to the rotation speed and a
high mass fraction compared to the stars. The morphology does not
depend on galaxy mass as much as it depends on evolutionary stage:
clump clusters are 100 times more massive than local dwarfs. The
apparent lack of star formation in damped Lyman alpha absorbers may
result from fast turbulence.
\end{abstract}


\section{Introduction}\label{intro}

Star-forming galaxies become increasingly irregular at higher redshift
with blue clumpy structure, asymmetry, and a lack of central
concentration (Glazebrook et al. 1995; Abraham, et al. 1996a,b; van den
Bergh et al. 1996; Driver et al. 1995, 1998; Im et al. 1999). These
three features are included in the CAS classification system (Conselice
2003), which has identified such irregularities in a large fraction of
galaxies in deep fields (e.g., Conselice, Blackburne, \& Papovich 2005;
Menanteau et al. 2006). Similar observations of irregular structures
were obtained from color dispersions in high redshift galaxies
(Papovich, et al. 2005), Gini coefficients (Lotz et al. 2006), Sersic
indices (Cassata et al. 2005; Ravindranath et al 2006), UV imaging (de
Mello et al. 2006), and various combined methods (e.g., Neichel et al.
2008).

The kinematics of disks at intermediate redshifts also show irregular
structures (Erb et al. 2004; Yang et al. 2008), suggesting an important
contribution from unstable gas dynamics (Law et al. 2009). Turbulent
motions are large compared to the rotation speed (F{\"o}rster-Schreiber
et al. 2006; Weiner et al. 2006; Genzel et al. 2006, 2008; Puech et al.
2007), although there can be underlying systematic rotation too (e.g.,
Bournaud et al. 2008).

During the last few years, we have examined the properties of clumps in
high redshift irregular galaxies, including chain galaxies (Cowie, Hu,
\& Songaila 1995) and their likely face-on counterparts, the
clump-clusters (Elmegreen, Elmegreen, \& Hirst 2004),  in an effort to
understand star formation and to look for signs of evolution toward
modern Hubble types. We have determined clump masses and ages from
population synthesis models and suggested the clumps form by
gravitational instabilities in a gas-rich disk (Elmegreen \& Elmegreen
2005; Bournaud, Elmegreen \& Elmegreen 2007, hereafter BEE). The clumps
are generally more massive than star complexes (Efremov 1995) in local
galaxies, which suggests that the turbulent speed in the neutral and
molecular gas is large as well, perhaps in the range of 20 to 50 km
s$^{-1}$, considering the characteristic mass of a disk instability
(and as measured in a z=1.6 clump cluster, Bournaud et al. 2008). Gas
column densities have to be large too, around $100\;M_\odot$ pc$^{-2}$
(Elmegreen, et al. 2009, hereafter EEFL). These properties are
reasonable considering the youthful stage of the systems we are
studying.  The high turbulence may come from gas accretion because it
has to be in place before the clumps form in order to define the clump
mass, and because the star formation feedback that generates turbulence
in local galaxies is relatively ineffective when the velocity
dispersion of the whole interstellar medium is large. High dispersion
star-forming clouds are tightly bound and not easily disrupted by
star-formation pressures (Elmegreen, Bournaud, \& Elmegreen 2008).
Clumps also produce high velocity dispersions by themselves in the
surrounding gas (BEE).

This interpretation of asymmetric clumpy structure as an indication of
instabilities in a gas-rich disk is not the only possibility. When
asymmetry and clumpiness are observed in a local L$^*$ galaxy, they are
usually indicative of a merger. Faint peripheral structures such as
tails and bridges contribute to this identification locally. As a
result, asymmetric and clumpy structure in high redshift galaxies has
also been considered to be the result of mergers (e.g., Conselice et
al. 2003, Treu et al. 2005; Lotz et al. 2008; Conselice, Yang, \& Bluck
2009, and references therein). This interpretation is reinforced by
expectations from the $\Lambda$CDM model, in which dark matter halos
grow by hierarchical merging (Davis et al. 1985). For example, a recent
study by Jogee et al. (2009) suggested that 16\% and 45\% of galaxies
with stellar masses larger than $2\times10^{10}\;M_\odot$ had a major
or minor merger, respectively, at redshifts between 0.24 and 0.80.
Jogee et al. identified these galaxies as mergers based only on their
asymmetric and clumpy structure, as determined both by eye and by the
CAS system. The actual fraction of galaxies that are clumpy in this
redshift range is smaller than the Jogee et al. merger fractions,
because they, like others, correct the observed fractions upward to
compensate for the low fraction of time during which a merger
morphology should be visible. L\'opez-Sanjuan et al. (2009) also used
the asymmetry index for galaxies in the range $0.35<z< 0.85$ and
derived a corrected major merger fraction of 20\%-35\% for $M_B<–20$
galaxies.

The key assumption for these and other merger interpretations is that
baryons come together in a clumpy fashion like the cold dark matter,
and star formation occurs early and efficiently in the baryonic clumps,
which then merge as little galaxies rather than as smooth gaseous
flows. Early numerical simulations reinforced this view, although the
results of these simulations depended strongly on the recipes for
thermal equilibrium and star formation, which are uncertain.
Significant merging is untenable in the instability model of clump
formation because then the pre-existing stars would make a spheroidal
component in the remnant (e.g., Abadi et al. 2003), and this component
would cause the instabilities to appear as spiral arms rather than
discrete clumps (Bournaud \& Elmegreen 2009).

Other aspects of galaxies expected from mergers are not generally
observed. Law et al. (2007) noted that the UV morphology of a galaxy is
not related to the star formation rate, which led them to conclude that
the irregular structure is probably not the result of a merger. Jogee
et al. (2009) also found that the star formation rate is not correlated
with galaxy morphology.  In local gas-rich mergers, even with weak
tidal forces, there is usually a significant increase in the star
formation rate compared to isolated galaxies (Larson \& Tinsley 1978;
see reviews in Sanders \& Mirabel 1996; Kennicutt 1998).

The thermodynamics of cosmic gas is the key issue in the theoretical
side of this debate.  Whether the gas, which tends to follow the dark
matter, cools enough to form stars in clumps before it assembles into
$M^*$ galaxies, or instead enters the $M^*$ halos in smooth flows,
depends on the balance between atomic collisional cooling and
compressional and radiative heating. Recent simulations that treat this
thermodynamics in detail now seriously question the baryonic merger
scenario. Murali et al. (2002) first did cosmological simulations with
enough resolution to include both large-scale flows and individual
galaxies. They found that cold flows of unprocessed gas can get
directly down to the $M^*$ scale without clumping into little galaxies
first. They suggested that galaxy growth is dominated by smooth flows
rather than mergers of pre-existing galaxies. Birnboim \& Dekel (2003)
confirmed in spherically symmetric collapse calculations that gas
cooling can be faster than compressional heating for low mass galaxies
(see also Binney 1977; Kay, et al. 2000), in which case the inflowing
material does not shock to the dark matter virial temperature. Semelin
\& Combes (2005) found the same dominance of cold cosmological inflow
to a disk, and noted that the final accretion tends to align to the
disk plane. Dekel \& Birnboim (2006) subsequently studied the stability
of halo shocks and showed simulations where cold gas streams penetrated
the hot halos. They found that the maximum halo mass for the cold flows
is comparable to the mass dividing blue and red galaxies in the local
universe and suggested that the difference between these two types of
galaxies is the result of a difference in the gas accretion mode.
Massive dark matter potentials shock their accreted gas to a high
temperature, which slows or prevents in situ star formation and tips
the balance of processes contributing to stellar growth in favor of
mergers. More recently, Dekel et al. (2009a) showed detailed
simulations and concluded that 2/3 of the inflow mass is smooth gas
accretion and the rest is clumpy merger-like accretion; they concluded
that most of the stars in the universe form in the disks of massive
($>10^{11}\;M_\odot$) ``stream-fed'' galaxies during the redshift
interval from 1.5 to 4. Dekel et al. (2009b) and Agertz, Teyssier, \&
Moore (2009) now find that cold flows can lead to the formation of
clumpy galaxies via gravitational instabilities in the accreted disks.

Following Murali et al. (2002), the same team now led by Kereš et al.
(2005) also did SPH simulations in a cosmological context. They showed
that accretion to low mass galaxies (baryonic mass
$<10^{10.3}\;M_\odot$; halo mass $<10^{11.4}\;M_\odot$) along
cosmological filaments remains cold and gets all the way to the central
disk, while high mass galaxies shock-heat the accreting gas before it
cools. They suggested that because of this mass dependence, the cold
mode dominates all galaxies at high redshift and is most important for
low density galaxies at low redshift. Recently, Kereš et al. (2009)
confirmed these results in a larger simulation and suggested that cold
flows dominate the star formation rate at all redshifts. In another
study, Ocvirk, Pichon, \& Teyssier (2008) included the effects of
metallicity. They derived the same threshold mass for virial shocks as
the other groups but suggested that the threshold mass for cold flow
penetration of hot halos increases with redshift as a result of changes
in the metal-dominated cooling rate.  At higher resolution, Brooks et
al. (2009) were able to study the time development of a galaxy disk
subject to shocked and unshocked inflows and to mergers of smaller
galaxies in a cosmological context. They found that unshocked gas
builds the disk much faster than shocked gas, which eventually accretes
slowly and for long times after cooling. In their model for a galaxy
the size of the Milky Way, the fraction of the disk stars coming from
merged galaxies is small, $\sim25$\%.

The distinction between clumpy disk structure that results from
gravitational instabilities in a highly turbulent interstellar medium
and clumpy disk structure that results from the merger of two or more
galaxies should be evident at moderate-to-low redshifts in the Great
Observatories Origins Deep Survey (GOODS; Giavalisco et al. 2004) and
Galaxy Evolution from Morphology and SEDs (GEMS; Rix et al. 2004)
fields. These surveys have exposure times that highlight the $z<1$
universe and are large enough to contain the relevant types as well as
rare intermediate cases. Galaxies with asymmetric clumps, galaxies with
double cores and tidal features (Elmegreen et al. 2007b), and galaxies
with smooth spiral arms, are all present in GEMS and GOODS over the
same redshift range. This mixture minimizes the bias from bandshifting
and surface brightness dimming.

With this in mind, we searched GEMS and GOODS for clump clusters, chain
galaxies, and spiral galaxies. For the clump clusters and face-on
spirals, we measured the magnitudes of star-forming clumps and their
adjacent interclump regions in the available ACS passbands. For the
chains and edge-on spirals, we measured the thicknesses of the disks.
There is generally an evolution toward smaller clumps and smoother
disks at lower redshifts. The relative number of combined clump
clusters and chains found in GOODS and GEMS is only $\sim10$\% compared
to spiral galaxies, while it is $\sim$50\% in the UDF out to $z\sim4$
(Elmegreen et al. 2005). We also found evidence for a progression in
relative clump mass, surface density, and age along the morphological
sequence from clumpy systems with no visible interclump stars, to
clumpy systems with red underlying disks, to spirals with relatively
smooth disks. This is the same evolutionary trend found in the UDF for
more distant galaxies (EEFL). The trend seems to illustrate how modern
disks and bulges form from the evolution, migration, and dispersal of
star-forming clumps.

The motivation in other studies to interpret high-redshift clumpy
asymmetric galaxies as merger remnants stems primarily from the
analogous morphology of local merger remnants, as discussed above.
However, there is another type of local galaxy with this morphology
that is not a merger remnant, the dwarf Irregular. We consider in
Sections \ref{sect:7793} and \ref{dwarf} the possibility that the
internal structure of high-redshift clumpy galaxies is a scaled-up
version of that in local dwarf Irregulars.  The local dwarfs get their
clumpy structure from large values of two dimensionless quantities: the
relative gas fraction in the disk and the relative length scale for
disk gravitational instabilities. The large unstable length compared to
the disk radius follows from another dimensionless quantity, the ratio
of the gas turbulent speed to the galaxy rotation speed. In the case of
dwarf Irregulars, the large value for this speed ratio is the result of
a low rotation speed ($50-100$ km s$^{-1}$) combined with a normal
turbulent speed ($\sim10$ km s$^{-1}$). If high redshift galaxies have
an equally large ratio, then it would arise from a high turbulent speed
at the normal rotation speed in an $M^*$ galaxy.

The disk instability model for clump formation also requires the gas
accretion rate to be larger than the star formation rate for at least
an orbit time. Otherwise, star formation would reduce the gas density
and make the layer more stable.  Such high temporary rates could be the
result of irregular inflows, where the gas and dark matter enter a
galaxy in separate bursts. Clumpy disks form or reform after the most
recent gas accretion event. Such a dependence between morphology and
accretion history may explain why clumpy disks exist over a wide range
of redshifts (Elmegreen et al. 2007a); i.e., clumpy structure at late
times could be initiated by a recent gas accretion event. Murali et al.
(2002) also consider late-time galaxy formation by recent cold flows.
Evidence for late stage galaxy formation was presented by Noeske et al.
(2007), based on the star formation rate versus mass for different
redshifts.

In what follows, the data used for the analysis of GOODS images are
described in Section \ref{data}, the clump masses, surface densities,
ages, and star formation rates are discussed in Section
\ref{properties}, and the disk thicknesses are in Section
\ref{thickness}. Section \ref{sect:7793} makes a comparison between two
clumpy, high-redshift galaxies and a local flocculent galaxy observed
with the same rest wavelength and convolved to the same spatial
resolution, and another similar comparison between a high redshift
galaxy and a local dwarf Irregular. The local flocculent is clearly
different from clump cluster galaxies in terms of the
clump-to-interclump mass contrast and brightness contrast, but the
local dwarf Irregular is indistinguishable except for a factor of $\sim
30$ in mass. A discussion of the implications of our study is in
Section 6: Section 6.1 reviews clump origins and trends with redshift,
Section 6.2 considers clump coalescence to make a bulge, Section 6.3
explores further the analogy with local dwarf Irregulars, and Section
6.4 offers a solution to the lack of star formation in damped Lyman
alpha absorbers. The conclusions are in Section 7.

\section{Data, Morphology, and General Implications of the Morphology}
\label{data}

The GOODS survey comprises images of 18 ACS fields surrounding the UDF
in 4 passbands, $B_{435}$, $V_{606}$, and $i_{775}$, and $z_{850}$ to
$z\sim1.5$. We searched all of these fields for clumpy galaxies of
various types and selected $\sim100$ for closer study. We also searched
5 GOODS fields for spiral galaxies and selected representative cases to
cover the same redshift range as the clumpy galaxies. Examination of
the clump cluster images suggested that some were questionable as
individual galaxies: some could be foreground-background pairs and
others could be interacting galaxies with tidal features. These were
rejected from the current study. Galaxies that were too highly inclined
to measure or distinguish the clumps were also rejected. This left a
sample of 93 galaxies: 26 spirals with clear spiral arms and bulges, 35
flocculent spirals with central red bulges, 15 clump clusters without
central red bulges but with an underlying red disk, and 17 clump
clusters with neither central red bulges nor obvious underlying disks.
Disk vertical scale heights were measured in an additional 62 chains
and edge-on spirals from GOODS.

The GEMS survey consists of 63 fields surrounding the GOODS fields to
slightly shallower depths ($z\sim1.2$) in 2 passbands ($V_{606}$ and
$z_{850}$). We selected a sample of 166 edge-on spirals and chain
galaxies larger than 10 pixels in diameter and measured their
perpendicular scale heights. We also selected a sample of 213 clump
clusters and measured 810 star-forming clumps. Because the accuracy of
the population synthesis fits is lower with only two passbands in GEMS,
their measurements were done as a check on the more detailed GOODS
measurements.

Figure 1 shows four morphologies of GOODS galaxies that are useful for
consideration here. On the left are two spiral galaxies that are
somewhat normal-looking compared to modern spirals. Next are two
galaxies with clumpy star formation and small red bulges. They resemble
local flocculent galaxies but have larger and fewer clumps than the
local analogs (see Section \ref{sect:7793}). The galaxies in the third
image from the left have clumpy star formation without an obvious
bulge, but there is still a red underlying disk in each. The two on the
right are clumpy without any obvious underlying disk. These four galaxy
types extend our previous classifications to more modern systems. In
Elmegreen et al. (2005), we classified disk galaxies in the Ultra Deep
Field (UDF) as either spirals or clump clusters, considering that a
third class, chain galaxies, represents an edge-on version of the clump
clusters. All but the leftmost pair in Figure 1 would have been called
clump clusters according to that classification, especially in the UDF
where the bulges in the two second-from-the-left galaxies would most
likely have been missed because of bandshifting and faintness. The
presence of bulges in some clump clusters was recognized in EEFL using
NICMOS images of the UDF. The three pairs on the right in Figure 1 have
classifications like galaxies in that EEFL paper: clump clusters with
bulge-like clumps, clump clusters with red disks and no bulge-like
clumps, and clump clusters without any evident red component. The first
of these, the flocculent class, has not been distinguished in our high
redshift surveys before. These are evidently normal disks that have
weak or no stellar density waves, like local galaxies with the same
appearance.

Figures \ref{fig:fig4-spirals-no-060409}-\ref{fig:Fig5-clumpy-060409}
show more examples of each morphology, presented in order of increasing
COMBO17 spectrophotometric redshift (Wolf et al. 2008).   The presence
of each type over a wide range of redshifts suggests that {\it clump
clusters are not bandshifted spiral galaxies}.  A similar redshift
comparison was made for 6 morphological classes in the UDF (Elmegreen
et al. 2007a). If the clumpy phase is short-lived, as simulations
suggest, then either {\it galaxy formation is prolonged} so that clumpy
galaxies appear even as late at $z\sim0.1$, or clump morphology is {\it
transient}, possibly following a significant event of gas accretion
late in the galaxy's life.

\section{Clump Properties}\label{properties}
\subsection{Method of Analysis} \label{method}

The magnitudes of 373 clumps were measured in four passbands for all of
the 93 selected galaxies in GOODS. Measurement was done using the
program $\it imstat$ in the Image Reduction and Analysis Facility
(IRAF\footnote{IRAF is distributed by the National Optical Astronomy
Observatories, which are operated by the Association of Universities
for Research in Astronomy, Inc., under cooperative agreement with the
National Science Foundation.}) with the same field position and size
for each passband (see discussion in EEFL). Clump boundaries were
typically $\sim10\sigma$ above the noise and measurement errors were
$\sim0.1$ mag. Boxes were used to define magnitudes because the clumps
are pixelated; a typical clump diameter is 3 to 5 pixels with a box
shape close to square. Clump color is much less variable than the clump
magnitude. Slight shifts in selecting the boundaries of these fields
would yield $\sim0.05$ magnitude deviations in the colors.

We measured the surface brightnesses of detectable interclump regions
that are adjacent to the clumps. This was done using the IRAF task $\it
pvector$, which takes a pixel-wide intensity cut through the galaxy,
and it was also done using selected rectangular regions with the IRAF
task $\it imstat$. Contours made with the IRAF task $\it contour$
provided further checks on the interclump brightness. The choice of
which interclump region to measure is subjective. We picked regions
fairly close to the clumps in most cases, and used the same interclump
measurements for several clumps if there were limited options. Clump
clusters are extremely clumpy and the surface brightness in the
interclump region varies a lot, from something that might be
representative of a clump pedestal to something too faint to detect at
all. Because we only measured regions considerably above the background
noise, there is a lower limit to the intrinsic interclump surface
brightness that increases as $(1+z)^4$, which is the cosmological
surface brightness dimming factor.

Figure \ref{fig:gems_hisdiff_bkgd_from_sky} shows histograms of the
difference between the $i_{775}$ surface brightness at a level of
$1\sigma$ noise in the sky and the $i_{775}$ surface brightness of each
interclump region. Solid blue lines are for spirals and dashed red
lines are for clumpy galaxies of various types. This figure indicates
that the average interclump region chosen for our study is about 2 mag
arcsec$^{-2}$ above the $1\sigma$ level of the sky, which is $25.2$ mag
arcsec$^{-2}$ in the $i_{775}$ band. This difference corresponds to a
factor of 6.3 above the $1\sigma$ noise level. The interclump surface
brightness is 2 to 3 times more accurate than $6.3\sigma$ because each
measured interclump region contains 4 to 10 pixels.  The peaked nature
of the distribution illustrates the point of the previous paragraph;
i.e., that most interclump measurements have about the same surface
brightness above the background noise level, in which case the
intrinsic interclump surface brightness, after correcting for
cosmological dimming, increases about as $(1+z)^4$.

The relative uncertainty per pixel in the surface brightness is
approximately the inverse square root of the counts. We noted above
that the clump boundaries are at about the $10\sigma$ level for
background noise $\sigma$, which makes the average count for the clumps
$\sim20\sigma$, and we also showed that the interclump regions are at
an average level of $6.3\sigma$. The ratio of these two mean
intensities is $\sim3$, and the inverse square root of this is
$\sim60$\%.

The flux from each star-forming region was determined by subtracting
the surface brightness of the adjacent interclump region from the
average surface brightness of the clump, and then multiplying the
result by the area of the clump (that is, the area of the box used to
determine the average surface brightness of the clump). Clump colors
were determined from the differences between the background-subtracted
clump magnitudes.  For the purpose of understanding clump dynamics, the
total clump mass, including the older stars inside the clump, should be
used, but for the purpose of understanding star formation, only the
excess young mass above the background should be used. Our previous
studies of clumps in UDF galaxies did not subtract the background disk
because it was generally very faint.

The observed colors $B_{435}-V_{606}$, $V_{606}-i_{775}$, and
$i_{775}-z_{850}$ of each background-subtracted clump and each
interclump region were fitted to three model parameters: age, star
formation decay time in an exponentially decaying model, and extinction
(see EEFL). We used the stellar evolution models of Bruzual \& Charlot
(2003) for the Chabrier initial mass function and a metallicity of
0.008 (equal to 0.4 solar). Estimates of dust absorption used the
wavelength dependence in Calzetti et al. (2000) with the
short-wavelength modification in Leitherer et al. (2002), considering
as templates six multiples ($M_A=0.25$, 0.5, 1, 2, 4, and 8) of the
redshift-dependent intrinsic $A_V$ in Rowan-Robinson (2003).
Corrections to the model spectra were made to account for intervening
hydrogen absorption, following Madau (1995). Spectrophotometric
redshift measurements come from the COMBO17 survey (Wolf et al. 2003;
2008). The templates considered decay times of
$\tau=10^7,\;3\times10^7,\;10^8,\; 3\times10^8,\;10^9,\;3\times10^9$
and $10^{10}$ years. Region ages were sampled at every timestep in the
Bruzual \& Charlot (2003) tabulation back to the age at a maximum
assumed region formation redshift equal to 10.  For each template, the
mass was determined from the age and the background-subtracted
brightness in the $i_{775}$ band.

The best fit results for age, decay time, extinction, and mass were
taken to be the $\exp\left(-0.5\chi^2\right)$-weighted average values
among all solutions, where $\chi^2$ is the sum of the squares of the
observed 3 ACS colors divided by the corresponding measurement errors.
Measurement errors were determined for each clump from the total count
of emission in that clump, after first scaling count rms to count value
using a large number of sample clumps (there is an approximately square
root relationship between the count rms and the count value).

Age and extinction are inversely correlated so each has a relatively
large uncertainty, but the effects of these two uncertainties
compensate for each other in the determination of mass, which is more
robust.  The mass errors resulting from uncertainties in metallicity
and extinction are relatively small and were discussed in detail in
EEFL. Here we estimate that the derived ages are uncertain ($3\sigma$)
to within a factor of 4 and the masses are uncertain to within a factor
of 2, based on 3 times the rms in the log of age and mass in the model
fits. Systematic uncertainties are larger and more difficult to
estimate, particularly regarding the star formation history, which is
not likely to be as simple as the exponential decay model assumed here.

A potential problem with fitting a number of parameters equal to the
number of measurements (three in our case) is that the most insensitive
parameter can vary evenly throughout the range and then the weighted
average value used for the fit is the average of the range.  We checked
for this by plotting versus age the restframe B-V color over the
redshift range from 0 to 0.65, which allows interpolation of the
observed magnitudes over the ACS bands to give the restframe apparent
magnitudes $m_B$ and $m_V$. Figure \ref{fig:gems_compare_age_color}
shows the result, with clump fits on the left and bulge or bulge-like
clump fits on the right. There is a correlation, indicating that the
fitting procedure is giving a sensible age that is younger for
intrinsically bluer regions. The scatter in the age is about half an
order of magnitude for the bluest colors and a few tenths of an order
of magnitude for the reddest colors. This half magnitude is consistent
with the factor of 4 for the $3\sigma$ age error mentioned above. Other
tests of the same fitting procedure were given in EEFL.

\subsection{Results}
\subsubsection{Clump Masses} \label{masses}

\begin{figure}\epsscale{1}
\plotone{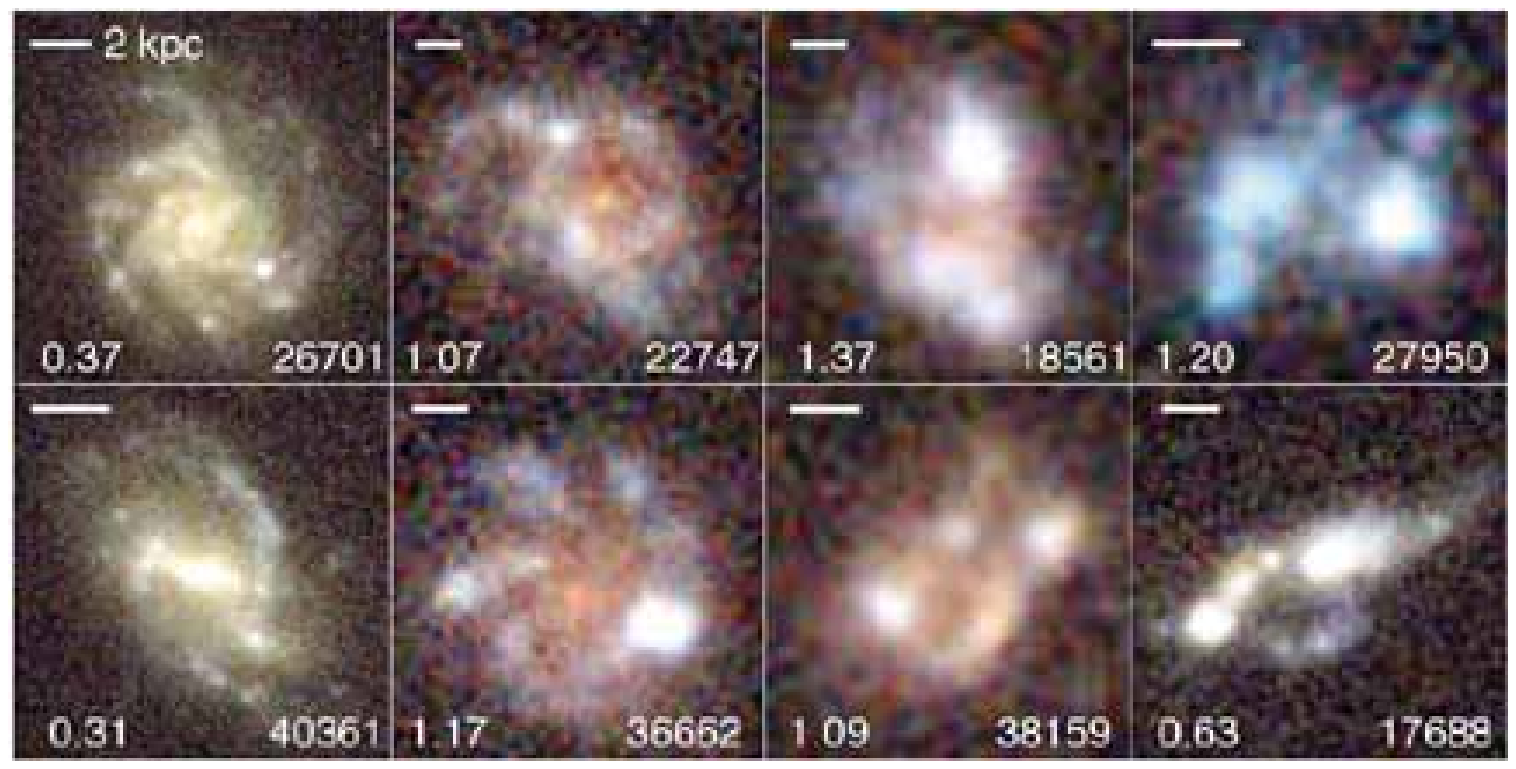} \caption{Color ACS Skywalker images of two spiral
galaxies on the left, flocculent spiral galaxies second left, clump
clusters with red underlying disks next, and clump clusters with no
evident underlying red disks on the right. Their Wolf et al. (2008)
redshifts and COMBO17 catalog identification numbers are given, along
with scale bars representing 2 kpc. [image degraded for arXiv
preprint]}\label{fig:GOODS-fig1-rev060409}\end{figure}

\begin{figure}\epsscale{1}
\plotone{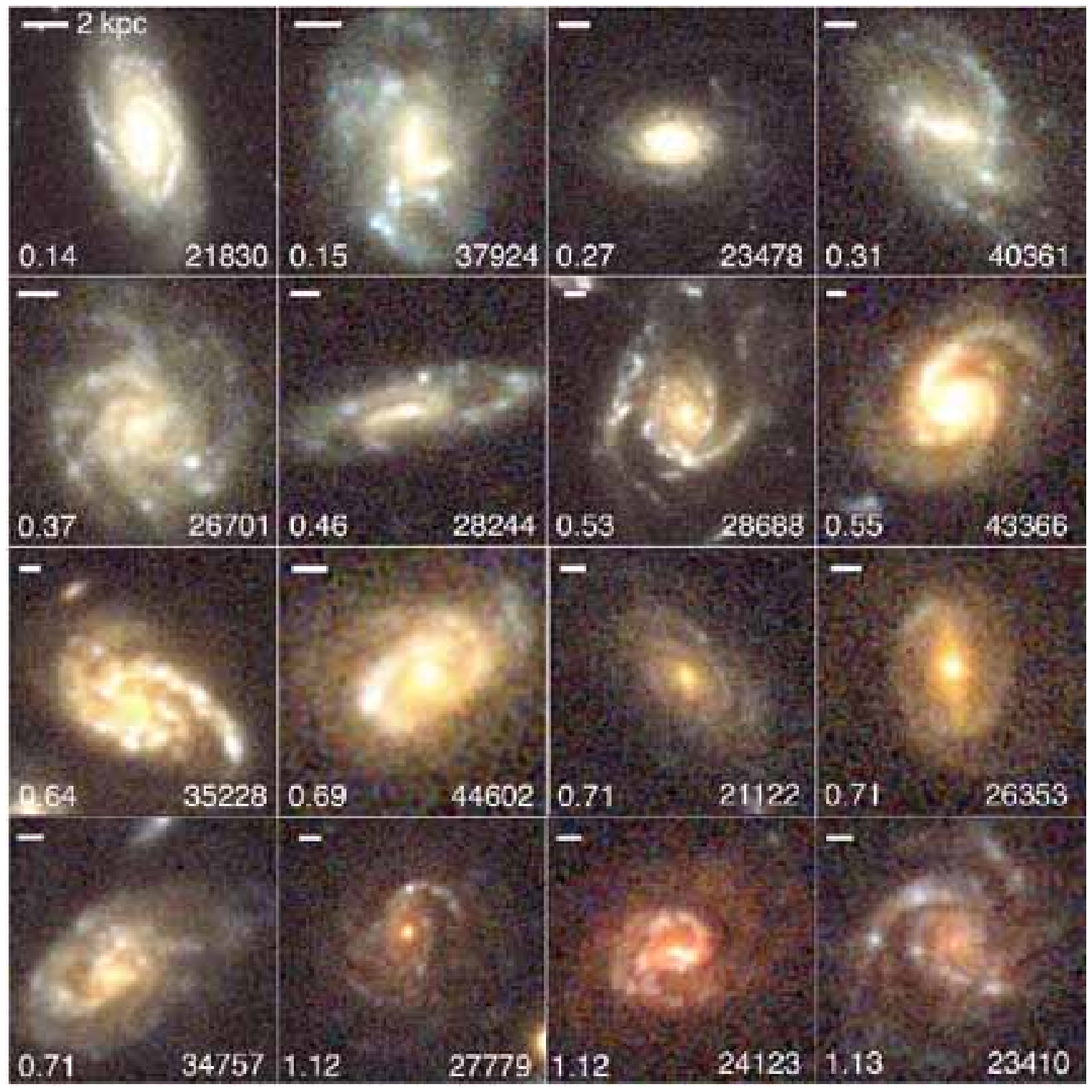} \caption{Color Skywalker images are shown for
spiral galaxies in the GOODS field. Their COMBO17 redshifts (lower
left), ID numbers (lower right), and 2 kpc scales (upper left) are
given. [image degraded for arXiv
preprint]}\label{fig:fig4-spirals-no-060409}\end{figure}

\begin{figure}\epsscale{1}
\plotone{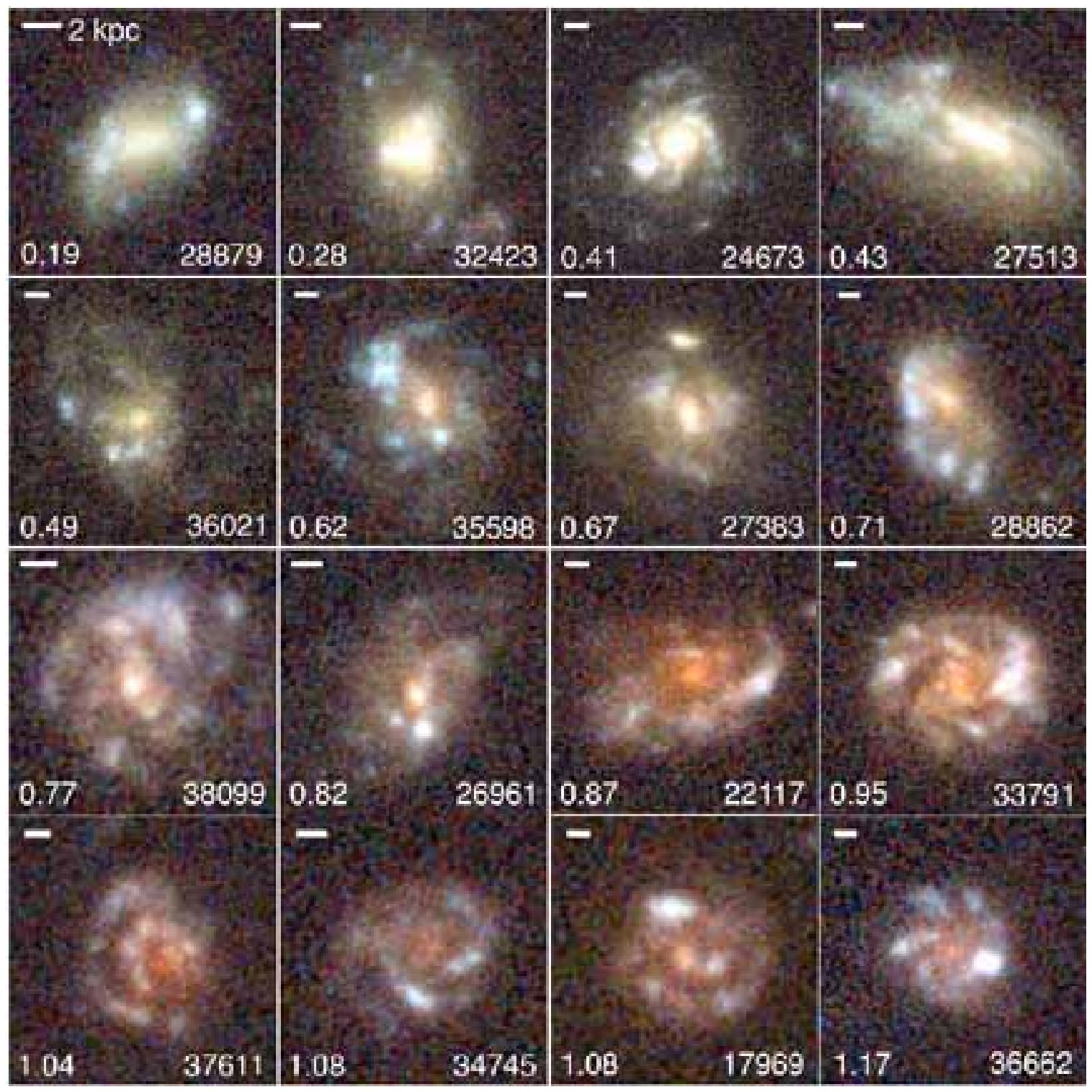} \caption{Color Skywalker images are shown for
flocculent spiral galaxies in the GOODS field. Their COMBO17 redshifts
(lower left), ID numbers (lower right), and 2 kpc scales (upper left)
are given. [image degraded for arXiv
preprint]}\label{fig:fig-flocclumpbulge-060409}\end{figure}

\begin{figure}\epsscale{1}
\plotone{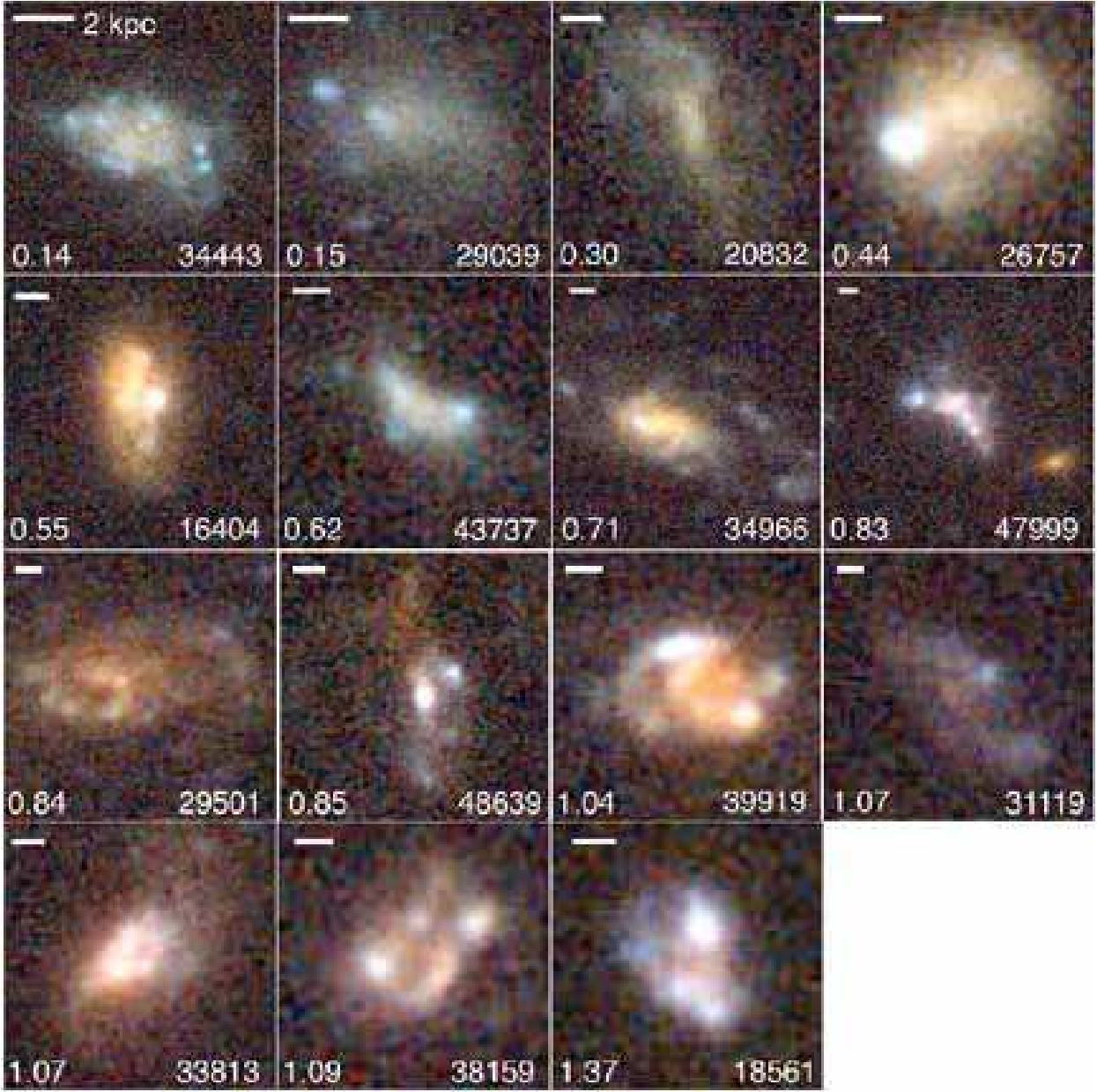} \caption{Color Skywalker images are shown for
clumpy galaxies in the GOODS field with red underlying disks. Their
COMBO17 redshifts (lower left), ID numbers (lower right), and 2 kpc
scales (upper left) are given. [image degraded for arXiv preprint]
}\label{fig:fig_clump-red-no-060409}\end{figure}

Figure \ref{fig:gems_masses} shows the best-fit masses for the clumps
versus the redshifts for galaxy types in the four divisions defined in
Figure \ref{fig:GOODS-fig1-rev060409}.  The bulges or bulge-like clumps
are indicated by open red squares and the clumps that are not
bulge-like are indicated by blue crosses.  The bulges are generally
more massive than the clumps; they are much more massive than the
clumps in the spiral and flocculent galaxies, while only a little more
massive than the clumps in the clump clusters. This is consistent with
our findings for the UDF discussed in EEFL. Beyond redshift $z\sim0.5$,
the average logarithms of the non-bulge clump masses (in $M_\odot$) for
spirals, flocculents, clump clusters with red disks, and clump clusters
without red disks are respectively,  $7.4\pm0.4$, $7.3\pm0.4$,
$7.4\pm0.5$, and $7.4\pm0.3$. The average log bulge masses at $z>0.5$
for the same galaxy types are $8.9\pm0.5$, $8.2\pm0.4$, $7.9\pm0.6$,
and $7.6\pm0.2$. Thus the bulges in spirals and flocculents are larger
than the clumps by an average factor of 16, while the bulges in the two
clumpy types are larger than the clumps by an average factor of only
2.2. This is consistent with UDF galaxies, where bulges are more
similar to clumps in the clumpiest galaxies than they are in the more
modern morphologies (EEFL).

The right-hand side of Figure \ref{fig:gems_masses} shows the ratios of
the clump or bulge masses to the whole-galaxy pseudo-luminosities,
measured as $10^{-0.4Brest}$ for restframe absolute magnitudes $Brest$
given in COMBO17 (Wolf et al. 2008). This ratio is convenient for
scaling the clump masses to the masses of star-forming regions in local
galaxies. The ratio is also useful for understanding a selection effect
evident in the left-hand panels; i.e., the drop in clump mass at lower
redshift.  This drop is not present in the right-hand panels,
indicating that the low clump masses at low redshift are the result of
systematically smaller and fainter galaxies at low redshift, with
essentially no change in the clump mass for a galaxy of a given
brightness.  The decrease in galaxy size for lower redshift is
presumably the result of a smaller sampling volume in the universe.

The log of the ratio of the clump mass to the galaxy pseudo-luminosity
averages $-1.1\pm0.5$, $-0.7\pm0.5$, $-0.4\pm0.5$, and $-0.1\pm0.4$ for
the spirals, flocculents, clump clusters with red disks, and clump
clusters, respectively. For the bulges, the logs of these ratios are
higher: $0.6\pm0.6$, $0.3\pm0.5$, $-0.1\pm0.6$, and $0.2\pm0.4$. These
averages consider all redshifts, so the differences between the logs
for the clumps and bulges in this case are slightly different than the
differences for the case of mass given above (which was for $z>0.5$).
The masses in clump clusters are larger than in spirals and
flocculents, relative to the galaxy luminosities, by a factor of
$10^{-0.1-[-1.1-0.7]/2}=6$.  Relative bulge masses compared to relative
clump masses are larger in spirals and flocculents than they are in
either type of clump cluster by a factor of $\sim11$, which is 10 to
the power $0.5(0.6+0.3-[-1.1-0.7])-0.5(-0.1+0.2-[-0.4-0.1])$.

In a typical local galaxy with $M_B\sim-20.3$ mag, the largest regions
of star formation contain $\sim10^5\;M_\odot$. In Figure
\ref{fig:gems_masses}, an average galaxy with $Brest=-20.3$ mag has a
clump mass of 10, 26, 52 and 105 million solar masses for the four
galaxy types. These are $\sim100$ to $\sim1000$ times larger than the
largest star-forming regions in local galaxies, and larger still if we
consider that the same galaxies in GOODS would be fainter today because
of stellar evolution. Locally, the value of $\log
\left(M/10^{-0.4M_B}\right)$ is $-3.1$ if $M=10^5\;M_\odot$ and
$M_B=-20.3$, much lower than the plotted values in Figure
\ref{fig:gems_masses}.

Some clumps in GOODS galaxies are 10 or more times larger than these
average values, considering the upper range of the points in Figure
\ref{fig:gems_masses}.  The maximum (non bulge-like) clump masses reach
$\log\left(M/10^{-0.4Brest}\right)\sim1$. At this value, a $M_B=-20.3$
mag galaxy would have a clump mass of $10^9\;M_\odot$. The trend toward
higher clump mass with redshift continues in the UDF. This trend
contains a selection effect determined by the observable limits of
surface brightness and physical size.  An important question is whether
the clumps define a physical scale, like a Jeans length, or a blending
scale at the limit of resolution in a distribution of smaller clumps.
We return to this question in Section \ref{GC}. Most likely, both
effects are present: the clumps probably contain unresolved
substructure, but the spacings between most giant  clumps are resolved
well enough to determine the clump luminosities and masses.

Figure \ref{fig:gems_masses_gems_galaxies} shows masses and
mass-to-light ratios for 810 clumps in 213 clump cluster galaxies in
the GEMS fields.  These masses were estimated from the $V_{606}$ and
$z_{850}$ filters used for GEMS in the manner described by Elmegreen et
al. (2007b). This method makes the same assumptions as in the rest of
the current paper, and uses the same Bruzual \& Charlot (2003) models
with exponentially decaying star formation rates. The method works
because for any given redshift, all of the ages and decay times in the
models give about the same track on a plot of clump apparent magnitude
$V_{606}$ versus color $V_{606}-z_{850}$ per unit stellar mass. These
tracks differ for each redshift, so we assign each galaxy to a redshift
interval of $\Delta z=\pm0.125$, where the tracks are determined. Given
the clump color on the abscissa of the plot, the deviation between the
clump apparent magnitude and the track apparent magnitude is
proportional to the log of the clump mass.  To find the best case, we
take the average in the log of clump mass from all of the different
tracks, which in fact have only small differences between them (see
Figs. 10 and 11 in Elmegreen et al. 2007b). The error in the mass is
estimated to be a factor of $\sim3$ from the relative deviations
between the tracks.  For the GEMS clump masses, we do not subtract the
underlying disk light in the $V_{606}$ and $z_{850}$ filters, as we do
for the GOODS masses. As a result, the GEMS masses tend to be larger
than the GOODS masses because the clump colors are slightly redder and
the clump fluxes are slightly larger without background subtraction.

Figure \ref{fig:gems_masses_gems_galaxies} shows that in GEMS also, the
clump masses increase with redshift yet have a nearly constant ratio of
mass to total galaxy light.  The average quantity
$\log\left(M/10^{-0.4Brest}\right)$ for GEMS clumps equals
$0.24\pm0.53$, which is larger than the equivalent quantity for GOODS
clump cluster clumps ($-0.1\pm0.4$) by 0.3 in the log. This difference
corresponds to a factor of $\sim2$ in mass, which is not unexpected
considering that the background is not subtracted for GEMS clumps.

\subsubsection{Clump Surface Densities}\label{surface}

The surface density of each clump was determined from its mass and
size. The size is the area of the region used to measure the clump flux
and is usually comparable to the size of the brightest part of the
clump. Because of angular resolution limits, each clump is likely to
have substructure where the surface density is larger than what we
derive here. Each clump has a nearby interclump region assigned to it,
but some clumps share the same region.  The surface density of each
interclump region was determined from its color and surface brightness
assuming the usual conversions between size and magnitude for a
$\Lambda$CDM cosmology (Carroll, Press \& Turner 1992; Spergel et al.
2003). Each fit to the interclump surface density included the mass,
age, exponential decay time, and extinction, as for the clump fits.

The left-hand side of Figure \ref{fig:gems_sb} shows the redshift
distribution of the excess surface density of each clump, written as
the total in each clump area minus the interclump surface density. The
units are $M_\odot$ pc$^{-2}$. Bulges have higher surface densities
than clumps by a factor of 10 to 100 for spirals and flocculents, but
the two are about the same for clump clusters. The right-hand part of
the figure shows the interclump surface density for the bulge and clump
regions (symbols as before). Both the clump and interclump surface
densities increase with redshift as $(1+z)^4$, which is the blue curve,
because of a detection limit: we can only measure detectable surface
densities (cf. Fig. \ref{fig:gems_hisdiff_bkgd_from_sky} and Sect.
\ref{method}) and these tend to be a constant value above the sky
noise.

The average vertical deviations between the points in Figure
\ref{fig:gems_sb} and the $\log(1+z)^4$ curves are a measure of the
relative surface densities for the four galaxy types. This quantity may
be written as  $\langle~\log\left(\left[\Sigma_{\rm clump} -\Sigma_{\rm
interclump}\right]/\left[1+z\right]^4\right)~\rangle$.  For non-bulge
clumps in spirals, flocculents, clump clusters with red disks, and
clump clusters, the average values of this quantity are $0.7\pm0.4$,
$0.4\pm0.4$, $0.6\pm0.4$, and $0.7\pm0.4$.  For the bulges, it is
$1.7\pm0.4$, $1.1\pm0.5$, $0.6\pm0.6$, and $0.7\pm0.4$, respectively.
Again we see that the excess surface densities are higher for bulges
than clumps, and they are higher yet for the spiral bulges and
flocculent bulges (average factor of 7) than for the clump cluster
bulges (average factor of 1). This latter result implies that if clump
clusters evolve into spirals, then the bulges in clump clusters have to
get denser with time (by an average factor of 6, which is 10 to the
power $0.5\left[1.7+1.1-0.6-0.7\right]$).

Similarly for the interclump regions, the averages of
$\log\left(\left[\Sigma_{\rm
interclump}\right]/\left(1+z\right)^4\right)$  for the four galaxy
types are $0.9\pm0.5$, $0.7\pm0.5$, $0.8\pm0.7$, and $0.3\pm0.6$,
respectively. Evidently, the interclump surface density is a factor of
$\sim3$ higher for spirals, flocculents, and clump clusters with red
disks than for clump clusters without red disks. This excess is
consistent with the conversion of clump cluster galaxies into spiral
galaxies as some fraction of the clump mass disperses into the
interclump medium.

Figure \ref{fig:gems_sb_withudf} shows the results again for the GOODS
clump clusters, but now including several higher redshift UDF galaxies
of the same morphological type.  The UDF galaxies are from Elmegreen \&
Elmegreen (2005), with photometric redshifts from Elmegreen et al.
(2007a) and analyzed in the same way as the GOODS galaxies. The minimum
detectable surface density continues to increase as $(1+z)^4$ because
of the brightness detection limit. For the interclump regions, the UDF
points are slightly below the curve while the GOODS points are slightly
above, reflecting the longer exposure time for the UDF.

The increase of intrinsic surface density with redshift is opposite to
the trend expected. Surface density should decrease with increasing
redshift as younger versions of galaxies are observed. Observations of
higher surface densities imply that we are seeing only the brightest
parts of the disks. At higher redshifts, these parts should come more
and more from the inner regions of the galaxies. Thus the trend of
increasing surface density should correspond to a trend of decreasing
observable galaxy size. Many studies have shown that galaxies appear to
be physically smaller at $z>1$ because we are observing primarily the
brighter inner regions of their disks (e.g., Buitrago et al. 2008;
Azzollini, et al. 2009 and references therein).

Figure \ref{fig:gems_ratio_histogram_comb} shows histograms of the
ratio of the clump star formation surface density to the interclump
surface density. Each count in the histogram is one clump. There is a
wide range in ratios, but generally the clump clusters have higher
ratios than the spirals and flocculents.  For clump clusters, the
typical star formation region has a mass surface density that is a
factor of $\sim2$ larger than the interclump surface density. This is
considered to be a lower limit to the true clump contrast for two
reasons: (1) the measured interclump surface density is always just
above the sky noise (cf. Fig. \ref{fig:gems_hisdiff_bkgd_from_sky}) and
therefore higher than the minimum interclump surface density in the
disk, which is probably unobservable; and (2) the clumps are barely
resolved and probably contain substructure or peaks with higher surface
densities. For spirals and flocculents, the clump/interclump contrast
is $\sim0.3$ with a wide range. Clump-cluster bulges have about the
same surface density contrast as the clumps, which is consistent with
their having comparable masses, shown above. Spiral and flocculent
bulges have much higher contrasts than the clumps, by factors of
$\sim3$ to $\sim30$.

A contrast of $\sim2$ between the surface density of the star formation
part of a clump and the surface density of the nearby interclump region
implies that the total clump/interclump surface density contrast, which
means the star formation plus the background in the clump, compared to
the background alone, is a factor of $2+1=3$. Evidently, the
star-forming clumps are significant gravitational perturbations in the
disks of clump clusters. This contrast is much larger than in local
galaxies. In the Milky Way, the average mass column density of a
molecular cloud is $\sim170\;M_\odot$ pc$^{-2}$ (Solomon et al 1987),
which is comparable to the stellar mass column density in the inner
regions of the disk. The star formation efficiency in such a cloud is
only a few percent, so the surface density of an OB association or star
complex is only a few percent of the background. As a result, it takes
$\sim100$ events of star formation in local molecular clouds to
significantly increase the surface density of a local stellar disk.
However, in clump clusters, a single event of star formation will
significantly increase the disk stellar surface density. For the clumps
to be so dense, the associated gas column density in the disk must be
comparable to or larger than the stellar surface density. Such high gas
mass fractions were also derived from the conditions required to make
the clumps by gravitational instabilities in the disk (BEE).

\subsubsection{Clump Ages}

Figure \ref{fig:gems_age} shows the ages in Gyr of the excess emission
from each clump (blue cross) and bulge (red square) versus redshift in
the left-hand panels, and the ages of the associated interclump regions
in the right-hand panels. The scatter in age is larger than the
uncertainty in each fit, which is a factor of $\sim4$ (Sect.
\ref{method}). This scatter is a result of continuous star formation in
these disks, so some regions are intrinsically younger than others.

Bulge ages are significantly older than clump ages for spirals and
flocculents, but about the same as clump ages for clump clusters. This
is consistent with our findings in EEFL. There is a slight trend toward
decreasing clump age with increasing redshift. We found this trend in
the UDF also (EEFL). In that previous paper, where the galaxies spanned
a wide range in redshifts, the age trend paralleled the age of the
local universe and so was partly a result of a real physical effect,
namely, that clumps and bulges in a young universe have to be young
themselves. In the present study, with a smaller redshift range, this
effect is expected to be smaller. There could also be some selection
effect involved because higher redshifts highlight bluer regions in the
disk, and bluer regions are younger.

Figure \ref{fig:gems_age_ratio_his_comb} shows histograms of the
logarithm of the ratio of the age of the excess emission from each
clump (indicated by the subscript ``clump-interclump'') to the age of
the associated interclump region. The histograms scatter around a ratio
of $\sim1$ ($\log\sim0$) for clump clusters with no red underlying
disks (bottom of the figure), and $\sim0.3$ ($\log\sim-0.5$) for clump
clusters with underlying red disks, spirals, and flocculents.

The age and surface brightness trends suggest an evolution from clump
clusters without red underlying disks to clump clusters with red
underlying disks, presumably as the clumps dissolve, age, and mix into
the disks. The trend continues to the spirals and flocculents.

\subsubsection{Clump Star Formation Rates}

Figure \ref{fig:gems_sfr} shows average clump star formation rates
determined from the ratio of each clump mass (above the background) to
its age (in $M_\odot\;{\rm yr}^{-1}$). The rates increase sharply with
redshift because of a combination of two selection effects: clump
masses increase with the galaxy detection limits, and clump ages
decrease because of an increasing bias toward the youngest components
of a clump at decreasing rest wavelength.  The increase in clump mass
with redshift is probably from a combination of effects: increased
clump blending as the physical resolution gets worse, an increased
Jeans length as the turbulent speed increases, an increase in absolute
clump surface brightness at the detection threshold, and an increase in
average galaxy luminosity with increasing cosmological volume. While
blending must be important for some considerations, blending does not
drive the increasing clump mass beyond $z\sim1.6$ (e.g., EEFL) because
the physical resolution begins to improve. Also for $z<1$, blending
does not cause the distinction between clump clusters and spiral
galaxies because both occur at the same redshift (Figs.
\ref{fig:fig4-spirals-no-060409}-\ref{fig:Fig5-clumpy-060409}).

Generally the clumps we measure are well separated so they are resolved
from each other. They also have a high contrast to the interclump
medium so we are not selecting marginally resolved local peaks in a
slowly varying background.  The fact that the ratio of the clump mass
to the total galaxy light is independent of redshift indicates that we
are not progressively smoothing over bigger and bigger subregions
within a galaxy as the spatial resolution worsens. More likely, most of
the clumps are identified correctly as discrete objects, and their
masses are measured correctly without severe blending effects, but the
whole galaxies are suffering a selection problem related to angular
resolution and surface brightness limitations. That is, we choose to
examine only galaxies that we resolve spatially (we limit our survey to
galaxies larger than 10 pixels in diameter) and that we see above the
sky noise surface brightness limit. These galaxy luminosities increase
with redshift by this selection effect (see Fig. 9 in EEFL), but for
each galaxy, the large clumps are distinct and the clump masses are not
themselves suffering an additional selection effect.

Fits to the redshift dependence of the star formation rate as
$\propto(1+z)^\alpha$ are shown in Figure \ref{fig:gems_sfr}, with blue
curves for the clumps and red curves for the bulges or bulge-like
clumps.  The average of all of the slopes $\alpha$ gives $SFR\propto
(1+z)^{8}$.  This sharp increase with redshift is the result of an
increase in clump mass and a decrease in clump age, both of which vary
by 1 or 2 orders of magnitude over the redshift range from 0 to 1. The
relation makes sense if we consider that the clump mass scales with the
galaxy luminosity (Fig. \ref{fig:gems_masses}) and the galaxy
luminosity scales with the limiting surface brightness multiplied by
the limiting resolved physical area, which is approximately a scaling
of $\propto(1+z)^4\times z^2$ for small redshifts. The clump age should
scale with the characteristic age of a star at the central restframe
wavelength for the ACS.  Stellar ages scale with their surface
temperatures $T$ as ${\rm age}\propto T^{-4}\propto (1+z)^{-4}$. Thus
the ratio of clump mass to age should scale with $\sim z^2(1+z)^{8}$ if
the observations are dominated by selection effects.  This is close to
what we see, which means that the individual clump masses, ages, and
formation rates are not characteristic of all star formation in these
galaxies (i.e., there are smaller and older clumps that are not
measured).

The right-hand side of Figure \ref{fig:gems_sfr} shows the product of
the clump age and the clump dynamical rate, $(G\rho)^{1/2}$, where
density $\rho$ comes from $3M/(4\pi R^3$) for clump mass $M$ and radius
$R=\left(M/\pi\Sigma_{\rm clump-interclump}\right)^{1/2}$.  This
product is approximately constant, or perhaps decreases slowly with
redshift, within the observed redshift range.  According to the
analysis in the previous paragraph, it should decrease approximately as
$(1+z)^{-2}z^{-1/2}$. The point-to-point scatter is much larger than
this factor.  Thus, selection effects are much smaller for this
dimensionless ratio than for the star formation rate itself. The
average clump age is about 1 dynamical time, with a scatter of a factor
of $\sim10$ either way.  Bulges are significantly older than clumps in
units of their dynamical time. This implies that star formation in
bulges has slowed down or stopped, and it also implies that bulges are
gravitationally bound.

The average value of unity for the product of age and dynamical rate is
reasonable considering that local star formation has about this same
value (Elmegreen 2007). However, another selection effect could be
present: fainter star-forming regions observed with the same physical
resolution limit would have lower densities and longer dynamical times.
If they are older, then they are redder and even fainter in restframe
blue passbands. Thus we tend to see the densest and youngest regions at
blue restframes. The youngest that any physically meaningful
star-forming region can be for its density is the dynamical age,
because this is how long it takes star formation to begin.  Thus the
value of unity is selected in any survey of the most easily observed
star-forming regions.

Figure \ref{fig:gems_sfr_udf} shows the average clump star formation
rate again for the clump clusters, but now with the UDF values added to
extend the redshift range. The product of the age and the dynamical
rate is on the right. The UDF points extend the trend seen in the
previous figure, considering that the spatial resolution scale stops
increasing and levels off at $z\sim1.6$.

\section{Disk Thickness}\label{thickness}

The intrinsic thicknesses of edge-on disk galaxies in the GEMS and
GOODS surveys were measured by fitting perpendicular profiles made with
the IRAF routine $\it pvector$ to Gaussian-blurred ${\rm
sech}^2\left(z/z_0\right)$ functions (see Elmegreen \& Elmegreen 2006).
The Gaussian blur accounts for the point spread function of the ACS.
Measurements of this function for ten stars in GOODS gave FWHM
dispersions of 3.21 px, 3.08 px, 2.87 px, and 3.18 px in B$_{435}$,
V$_{606}$, i$_{775}$, and z$_{850}$ filters, respectively. (Note a
typographical error in Elmegreen \& Elmegreen [2006] where we quote a
Gaussian sigma for the ACS stellar images of about 3 px but actually
mean and use a FWHM equal to this value.) At redshifts of 0.1, 0.3, and
1, a FWHM of 3 pixels corresponds to a projected distance of 160, 400,
and 720 pc. At higher redshift in the UDF, the projected distance gets
slightly smaller; e.g., it is 640 pc at $z=4$. For chain galaxies, the
perpendicular profiles were taken to be as wide in the parallel-to-disk
direction as the major axes, to minimize the pixel noise. For edge-on
spirals (distinguished by their bulges in the ACS images), two wide
profiles were taken, one on each side of the bulge, and then averaged.
All ACS passbands were measured and fit, but here we discuss only the
fit to the observed $V_{606}$-band image. There is a slight increase in
disk thickness with wavelength (see Elmegreen \& Elmegreen 2006).

The top right panel of Figure \ref{fig:findHandMB} shows the resultant
thicknesses $z_0$ versus the absolute restframe B-band magnitudes, from
COMBO17 (Wolf et al. 2008). Each symbol represents a GEMS or GOODS
edge-on galaxy. Spirals (plus symbols) and chains (dots) have about the
same thicknesses (as in the UDF; Elmegreen \& Elmegreen 2006). The
other panels in Figure \ref{fig:findHandMB} show: local galaxies in the
top left, UDF chains in the lower left, and UDF spirals in the lower
right. Each panel has a solid line showing the indicated linear fit to
the points in that panel, and three dashed lines showing the linear
fits to the points in the other panels (color coded), for comparison.
The UDF results are from our previous paper; the dots with circles
represent the best cases for measurement (the linear fits include all
galaxies plotted). The local scale heights were determined from a ${\rm
sech}^2$ fit to R-band images; plus symbols are from Yoachim \&
Dalcanton (2006), $x-$symbols are from Barteldrees \& Dettmar (1994),
and open circles are from Bizyaev \& Kajsin (2004).

The left-hand panel of Figure \ref{fig:findHandMB_vs_z_corrected_only}
shows the scale height $z_0$ versus redshift for GEMS and GOODS spirals
(plus symbols) and chain galaxies (dots) and UDF chain galaxies
(crosses). The two groups have similar dependencies. There is a
decrease in $z_0$ for low redshift, as there was a decrease in clump
mass for low redshift in Figure \ref{fig:gems_masses}. Both decreases
arise because the galaxies in these surveys are intrinsically fainter
at lower redshifts. The green curve shows the FWHM of point sources in
the GOODS images, taken to be a constant 3.0 px.  The lower envelope of
the point distribution is about the FWHM, so the thinnest disks are
barely resolved.  The right-hand panel of Figure
\ref{fig:findHandMB_vs_z_corrected_only} shows the difference between
the measured scale height and the scale height at the restframe $M_B$
of the galaxy that comes from the linear fit in Figure
\ref{fig:findHandMB}.  This $M_B-$corrected scale height has no
redshift dependence and is the same for GEMS, GOODS, and UDF chains.

According to the linear fits in Figure \ref{fig:findHandMB}, the scale
height of an $M_B=-20.3$ mag galaxy is $\sim1.2$ kpc locally, $\sim1.2$
kpc in GEMS and GOODS, $\sim0.86$ kpc for UDF chains, and $\sim1.0$ kpc
for UDF spirals, with $\sim30$\% variations around these values. These
scale heights are all about the same at this absolute magnitude.
Galaxies tend to be thinner at fainter magnitudes, and local faint
galaxies appear thinner than high redshift faint galaxies by about
30\%. This difference is too small to be significant considering the
relatively poor angular resolution of the high redshift disks.

Figures \ref{fig:findHandMB} and
\ref{fig:findHandMB_vs_z_corrected_only} suggest that clumpy galaxies
and spiral galaxies in GEMS and GOODS have about the same thickness
when viewed edge-on. This is also about the same as the thickness of
galaxies in the local universe when scaled to the same absolute
restframe blue magnitude. High redshift galaxies should fade over time,
however, and the thickness of the parts currently observed at high
redshift could change as well, with disk accretion increasing the
gravitational force and causing a shrinkage, and satellite accretion or
stellar scattering off clouds and spiral waves stirring the disk and
causing an expansion.

To estimate fading over time, we use the population evolution models in
Bruzual \& Charlot (2003) for a Chabrier IMF and a metallicity of 0.4
solar (as elsewhere in this paper).  In one of their tables, the
absolute B-band magnitude per unit solar mass of stars varies with age
$T$ in Gyrs approximately as $M_B=4.88+2.37\log(T)$ mag. Then a change
in $T$ from 1 Gyr to 5 Gyr corresponds to an increase in $M_B$ by 1.66
mag; a change in $T$ from 3 Gyr to 10 Gyr increases $M_B$ by 1.24 mag.
If we consider these values to be typical and take 1.5 mag of fading
for this population in the restframe B band, and if we combine this
fading magnitude with the fitted relation in Figure
\ref{fig:findHandMB}, $\log z_0=-1.312-0.067M_B$, then the thickness
ends up too large for its faded magnitude by $0.067\times1.5=0.10$ in
the log, or a factor of 1.26 in $z_0$. This argument suggests a way in
which old components of today's disks, viewed directly in GEMS and
GOODS, can end up thicker than the young components by the time they
are viewed in a modern galaxy. Satellite stirring and stellar
scattering in the disk would do the same thickening with age, but here
we see the thicker components of the main disk before subsequent
kinematical evolution. We do not believe we are seeing what is called a
thick-disk component, however. That would be thicker than our observed
values by a factor of 2 for a given $M_B$, considering the thick disk
measurements in Yoachim \& Dalcanton (2006).

To check disk thickness in another way, we measured the major and minor
axes lengths from IRAF {\it pvector} scans in the $V_{606}$ band for 46
chain galaxies and clumpy edge-on spirals in GOODS that had reliable
redshifts in Wolf et al. (2008).  The perpendicular scans used for this
were the same wide scans used to determine $z_0$; these scans are as
wide as the galaxies are long. For the spirals, there are two scans,
one on each side of the bulge, and the average of the two widths was
used. The parallel scans are as wide as the galaxies are thick. The
axes endpoints were determined at intensities equal to half of the
local peak. For a minor axis, this was generally half of the total
peak, but for a major axis, this was half of the peak intensity of the
part at each end, even if there was a brighter part in the center. The
point of this procedure is that it allows us to subtract the FWHM of a
stellar image from the measured axis length in quadrature, to correct
for the instrument point spread function. We take a FWHM of 3.08 px
from Gaussian fits to stars in the $V_{606}$ band.  The average minor
axis width after correction for the point spread function is
$8.1\pm2.3$ px, which is enough larger than the FWHM of a star for us
to be confident that we have resolved this length. The average ratio of
minor to major axis is $0.16\pm0.06$. The range of ratios is 0.06 to
0.33. There was no significant difference between the chains and the
spirals in these ratios.  This value of 0.16 is somewhat large compared
to local edge-on, late-type spiral galaxies, where a typical minimum
ratio for edge-on cases is $\sim0.1$ (from the de Vaucouleurs et al.
1991 atlas; see Figure 9 in Elmegreen et al. 2005). It is larger still
compared to the flattest local galaxies (Karachentsev et al. 2000). The
difference between the GOODS and local galaxies is not considered to be
significant, however, given the poor resolution in GOODS and the
unknown inclinations of clumpy systems.

The observed axial ratios for the outer isophotes of these edge-on
galaxies are larger without the correction for instrument point spread
function. At the level of $2\sigma$ sky noise, the average axial ratio
for the same galaxies is $0.270\pm0.078$.

In another test of disk thickness, we measured the radial exponential
scale lengths, $h_R$, of most of the edge-on spirals in GEMS and GOODS
that are larger than 10 pixels, which is 113 galaxies.  This was done
in all ACS passbands, but we discuss the $V_{606}$ measurements here.
These lengths were determined from thick parallel scans using {\it
pvector} as above, and fit to a Gaussian-blurred model of an
exponential disk viewed edge on. The Gaussian blur corrected for the
point spread function, taken to be 3.08 px from stellar images. The
halfwidths $z_0$ were also determined for these spiral galaxies by
fitting to a Gaussian-blurred ${\rm sech}^2(z/z_0)$ function, as above.
Then we determined the ratio of the scale height to the scale length,
$z_0/h_R$. The average value was $0.43\pm0.20$. Comparing this to the
ratio for local galaxies in Figure 5 of Yoachim \& Dalcanton (2006), we
see that the ratio in GOODS is larger than the ratio for local spirals
by a factor of 2 to 3. It is larger even than the ratio for local
galaxies with low circular velocities (dwarf Irregulars), which has the
largest ratio among local types, equal to $\sim0.2$ on average. This
result does not mean that the GOODS galaxies are particularly thick,
however, as there should be extinction corrections from dust in the
midplane. The average ratio is therefore viewed to be unreliable.

The above three paragraphs suggest that the disks in GEMS and GOODS
could be slightly thicker for their magnitudes or lengths than the
disks in local spirals.  We are highly constrained by the available
resolution of the ACS, however. We find, for example, that the physical
lengths of both the minor and major axes (in kpc) increase with
redshift (as on the left in Figure
\ref{fig:findHandMB_vs_z_corrected_only}). Presumably the thinnest
disks at high redshift are too faint to include in our survey.

The scale height from Figure \ref{fig:findHandMB_vs_z_corrected_only}
and the mass column density of the interclump medium from Figure
\ref{fig:gems_sb} can be combined to give a velocity dispersion,
$\sigma$. Using the thin disk formula, $\sigma^2=\pi G\Sigma_{total}
z_0$, we get $\sigma=20\left(\Sigma_{total}/30\;M_\odot\;{\rm
pc}^{-2}\right)^{1/2}\left(z_0/kpc\right)^{1/2}$ km s$^{-1}$. This
normalization value of $\Sigma_{total}=30\;M_\odot\;{\rm pc}^{-2}$ is
comparable to the stellar value $\Sigma_{\rm interclump}$ in Figure
\ref{fig:gems_sb}. If there is a significant amount of gas, then the
total column density would be larger, possibly making $\sigma\sim30$ km
s$^{-1}$ or more. This is comparable to the velocity dispersion of
stars in the solar neighborhood.

\section{A Comparison between GOODS Galaxies, a Local
Flocculent Galaxy, and a Local Dwarf Irregular Galaxy}\label{sect:7793}

The clumpy appearance of some galaxies in this study is reminiscent of
that in local flocculent galaxies. Figure
\ref{fig:fig19-N7793,GOODS-060409} shows a comparison between two GOODS
galaxies and the local flocculent NGC 7793, blurred to the same spatial
resolution and viewed at the same restframe wavelength. The top left
panel shows a IIIa-J (3950 \AA) image of NGC 7793 taken with the UK
Schmidt telescope and obtained from the Digital Sky Survey at the Space
Telescope Science Institute (MAST). The top right panel shows the
$B_{435}$ image of the GOODS galaxy 34443, which has a redshift of
$z=0.139$ (Wolf et al. 2008). The restframe wavelength for this galaxy
is $4350/1.139=3819$\AA, close to the wavelength of the NGC 7793 image
to its left.

The blurring of NGC 7793 was done as follows. For $z=0.139$ in the
GOODS galaxy, one pixel in the ACS camera, which is 0.03'', corresponds
to a projected spatial size of 72.6 pc. The average FWHM of stars in
the ACS image at $B_{435}$ band was measured to be 3.2 pixels, so the
FWHM of point sources appears to have a size of 230 pc in the image of
34443.  To make a blurred image of NGC 7793 with the same physical
scale for the FWHM of a point source, we first note that the original
image scale is 1.7'' per pixel, and the average FWHM of several stars
in the field is 2.05 pixels. NGC 7793 is at a distance of 3.1 Mpc
(NASA/IPAC Extragalactic Database), and at this distance, the desired
FWHM resolution scale of 230 pc subtends 15.5,'' which is 9.1 px.  Thus
we blur the original image of NGC 7793 with the routine {\it Gauss} in
IRAF using a Gaussian convolution function with a Gaussian sigma that
produces a net FWHM of 9.1 px. Considering that the original image has
a FWHM of 2.05 px, this means we have to blur it with an additional
FWHM of $\left(9.1^2-2.05^2\right)^{1/2}=8.87$ pixels. The
corresponding Gaussian sigma for the blur is
$8.87/\left(8\ln2\right)=3.77$ pixels.

The image of NGC 7793 is further degraded to make the pixel scale and
the noise level about the same as for 34443. As mentioned above, the
number of pixels in a resolution FWHM for the blurred image of NGC 7793
is 9.1 px, and the number in 34443 is 3.2 px. The ratio of these is
2.8, so we re-pixelate the blurred NGC 7793 image by converting each
block of $3\times3$ pixels into a single pixel using the {\it blkavg}
routine in IRAF.  Also, the ratio of the number of counts in the peaks
of 34443 to the rms of the sky was measured to be about 10, so we
subtracted sky from the blurred re-pixelated image of NGC 7793 and
added noise using the IRAF routine {\it mknoise}, giving it the same
ratio of peak intensity to sky rms.  The result of all of these steps
is an image of the local flocculent galaxy NGC 7793 that has the same
rest wavelength, physical resolution, pixelation, and noise level as
the GOODS galaxy 34443. Figure \ref{fig:fig19-N7793,GOODS-060409} shows
the two images with about the same linear and angular scales.

The bottom two panels in Figure \ref{fig:fig19-N7793,GOODS-060409} make
a similar comparison between NGC 7793 and the GOODS galaxy COMBO17
17969 at $z=1.08.$ In this case, the image on the left is the NUV (2267
\AA) image of NGC 7793 from the {\it Galaxy Evolution Explorer}
satellite ({\it GALEX}; Martin et al. 2005). It has an image scale of
1.5'' per pixel and a FWHM measured for several point sources of
$\sim3.38$ px. The image on the right is the $B_{435}$ image of 17969,
which has a restframe wavelength of $4350/2.08=2090$\AA, close to that
of the NGC 7793 image. At $z=1.08$, one px in the ACS corresponds to
245 pc, so the FWHM of a point source in 17969 has a spatial scale of
790 pc. We want to blur the NGC 7793 image to the same physical scale.
At a distance of 3.1 Mpc, 790 pc subtends an angle of 52.7,'' which is
35.1 px.  The intrinsic FWHM of the GALEX image is 3.38 px, so we have
to blur it with an additional $\left(35.1^2-3.38^2\right)^{1/2}=35.0$
pixels FWHM. Converting this to a Gaussian, we get $\sigma=14.9$ px for
the IRAF routine {\it Gauss}.  Then the physical resolution of the NGC
7793 and 17969 images are the same (790 pc). Next we re-pixelate NGC
7793 using the routine {\it blkavg} with a box size equal to the ratio
of the FWHM of a point source in NGC 7793, 35.1 px, to the FWHM of a
point source in 17969, 3.2 px; the closest integer to this ratio is 11.
Finally we subtract sky and add noise to the NGC 7793 image.  The ratio
of the intensity of a typical peak in 17969 to the rms of the sky is
$\sim20$, so we add noise to the degraded image of NGC 7793 to give
this same ratio.  The result is the image of NGC 7793 in the lower left
of Figure \ref{fig:fig19-N7793,GOODS-060409}, with the same rest
wavelength, physical resolution, pixelation, noise level, and scale as
the GOODS galaxy 17969.

The blurred images of NGC 7793 have blended star formation regions that
are about the same diameter as the star formation regions in the GOODS
galaxies. NGC 7793 has a prominent exponential disk, however, so the
central region looks like a big clump. This is not the case for the
clump clusters. Also, in the bottom left of Figure
\ref{fig:fig19-N7793,GOODS-060409}, the two biggest clumps in NGC 7793
look like projection-enhanced parts of the exponential disk because
they are on the minor axis. If such disks are also present in high
redshift clumpy galaxies, then they have to be much fainter relative to
the clumps than in local galaxies in order to have the high
clump/interclump contrast shown in Figures
\ref{fig:gems_ratio_histogram_comb} and
\ref{fig:fig19-N7793,GOODS-060409}. For GOODS galaxy COMBO17 17969 in
the bottom right of Figure \ref{fig:fig19-N7793,GOODS-060409}, the
clumps stand out sharply from the rest of the disk in the restframe UV,
much more than the clumps in NGC 7793 at the same NUV wavelength.  Thus
a primary difference between the GOODS clump clusters and a local
flocculent galaxy is the high surface density contrast of the clumps in
the GOODS sample.  We made the same point in Section \ref{surface}.
Other differences are the small number and high mass of distinct clumps
in the GOODS galaxies compared with local spirals.

Figure \ref{fig:HoII,18561-060409} shows a comparison between a GOODS
clump cluster and a local dwarf Irregular, Ho II, at a distance of 3.48
Mpc (NASA/IPAC Extragalactic Database). On the left is the NUV image
(2267 \AA) of Ho II at full resolution, which is 1.5'' per pixel with a
FWHM of 3.4 px for a point source. In the center is Ho II blurred with
a Gaussian $\sigma=13.1$ px so that the FWHM of a point source has a
size of 780 pc. On the right is the $V_{606}$ ACS image of COMBO17
18561, which has a photometric redshift of 1.367 (Wolf et al. 2008).
The rest wavelength is $6060/2.367=2560$\AA, about the same as in the
Ho II image. The FWHM of a point source is 3.08 px, with a 0.03''
pixel. At its redshift, this FWHM corresponds to 780 pc, the same FWHM
as for Ho II. Thus the GOODS image of 18561 on the right has the same
resolution and restframe wavelength as the blurred NUV image of Ho II
in the center. The two also have about the same pixel scale, relative
noise level, and page scale, as discussed for Figure
\ref{fig:fig19-N7793,GOODS-060409}. Evidently the degraded local dwarf
Irregular in Figure \ref{fig:HoII,18561-060409} looks qualitatively
similar to a clump cluster, unlike the flocculent spiral in Figure
\ref{fig:fig19-N7793,GOODS-060409}.

Quantitatively, Ho II and COMBO17 18561 are very different, however.
The apparent B magnitude of Ho II is 11.13 (Bureau \& Carignan 2002)
and the distance is 3.48 Mpc, so the absolute B magnitude is $-16.6$.
According to Wolf et al. (2008), the restframe absolute B magnitude of
18561 is $-20.37$, a factor of $\sim30$ brighter. The prominent clump
in the upper part of 18561 has an apparent $z_{850}$ magnitude of 25.4
without background subtraction. This passband corresponds to a
restframe wavelength of 3600 \AA, similar to U-band. For the redshift
of 1.367, the distance modulus is 44.96 so the absolute restframe U
magnitude of the clump is $-19.6$. The U-B color of Ho II is -0.1
(Stewart et al. 2000) so the U-band magnitude of Ho II is $-16.7$. Thus
the prominent clump in 18561 is 2.9 magnitudes, or a factor of 14,
brighter than all of Ho II in the restframe U band. The mass we derive
for the star-forming part of this clump is $1.3\times10^8\;M_\odot$ and
the age we get is $\sim3$ Myr. This is the same age as the central
clump in Ho II, which contains $\sim170$ O stars and is $\sim100$ pc
across (Stewart et al. 2000).

\begin{figure}\epsscale{1}
\plotone{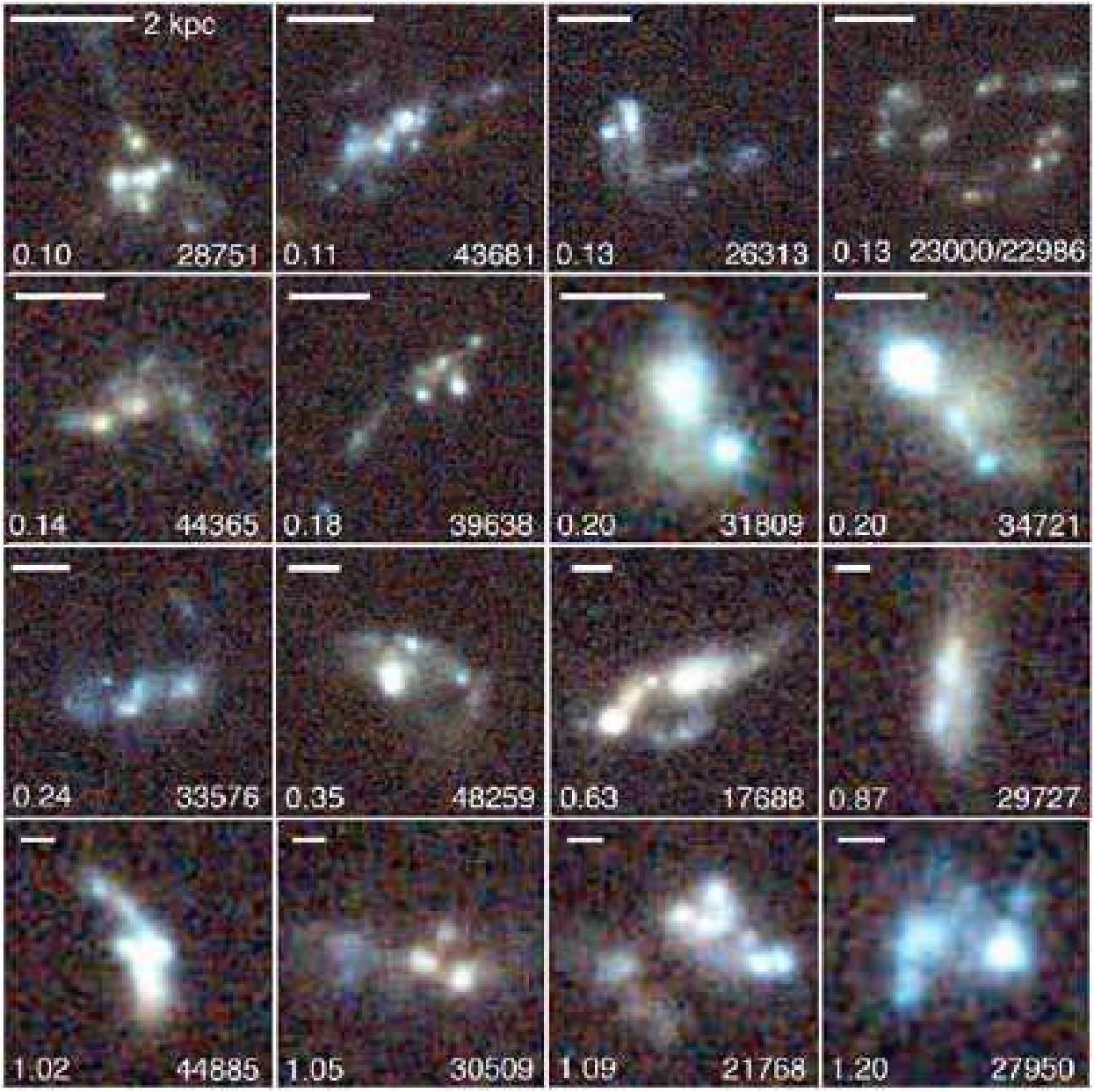} \caption{Color Skywalker images are shown for clump
clusters in the GOODS field. Their COMBO17 redshifts (lower left), ID
numbers (lower right), and 2 kpc scales (upper left) are given. Their
appearance is dominated by blue clumps, with no evidence for a red
underlying disk. Some could be mergers; others are probably in-situ
disk star formation. [image degraded for arXiv
preprint]}\label{fig:Fig5-clumpy-060409}\end{figure}

\begin{figure}\epsscale{1}
\plotone{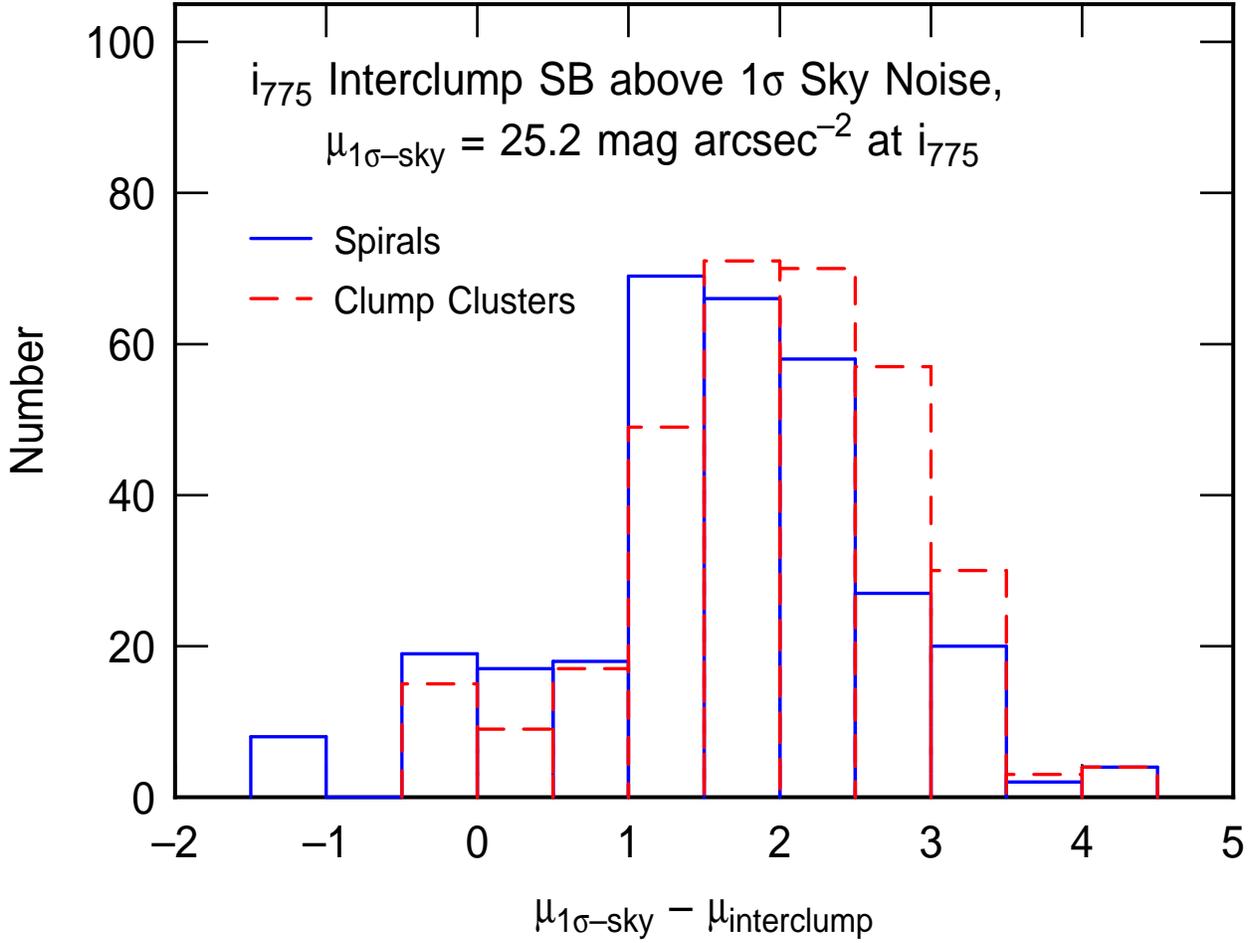} \caption{Histograms of the differences between the
surface brightness equivalent to $1\sigma$ sky noise and the surface
brightness of each interclump region. Interclump regions were chosen
close to the clumps. The typical interclump region is about 2 mag
arcsec$^{-2}$ above the sky noise.  The value at $-1.25$ mag
arcsec$^{-2}$ corresponds to a region near a clump where the image
count was negative, meaning it was fainter than the average sky value.
}\label{fig:gems_hisdiff_bkgd_from_sky}\end{figure}

\begin{figure}\epsscale{1}
\plotone{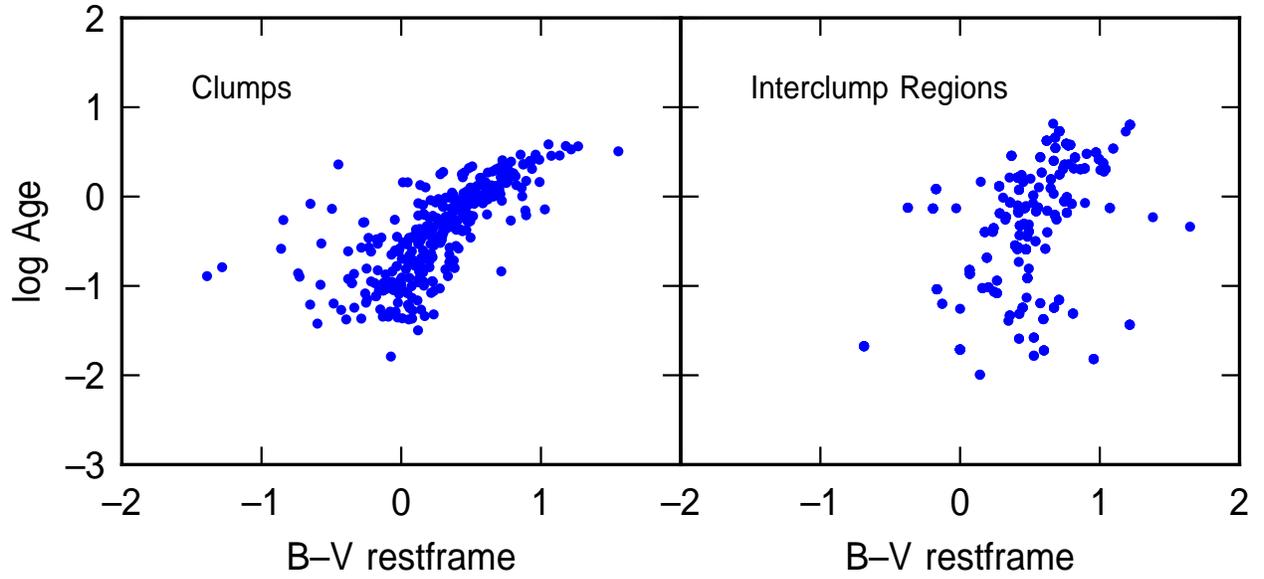} \caption{As a check on the model fits, we show the
expected correlation between restframe color and clump age (on the
left) and interclump age (on the right).  The ages are in Gyr. The
fitting procedure is giving a sensible age that is younger for
intrinsically bluer
regions.}\label{fig:gems_compare_age_color}\end{figure}

\begin{figure}\epsscale{.7}
\plotone{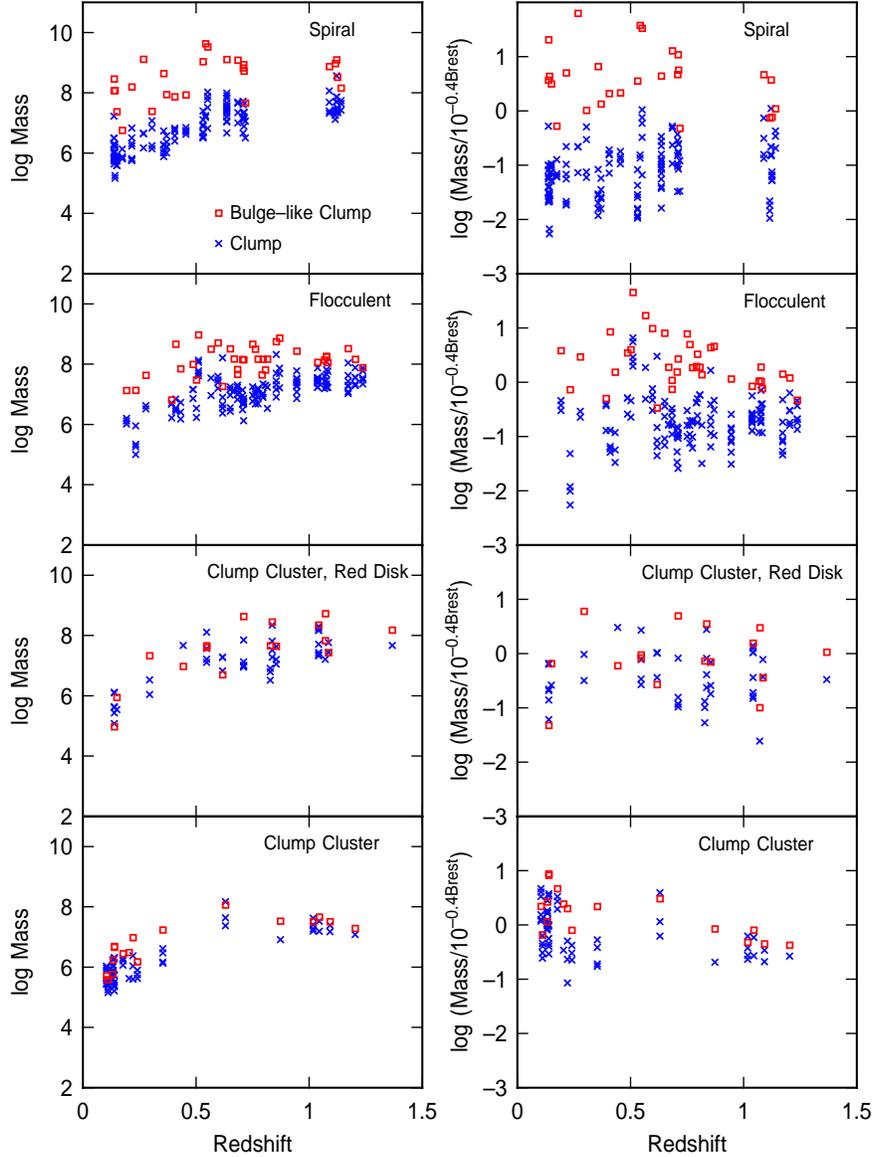} \caption{Clump mass (left) and clump mass per unit
galaxy light (right) are shown versus redshift. Red squares are for
bulges in the case of spirals and flocculents, and bulge-like, or
centralized clumps in the case of clump clusters.  Blue crosses are for
clumps outside the bulge regions. All mass determinations have had the
underlying disk light removed using surface brightness measurements of
adjacent regions. Bulges are more massive than clumps by a factor of
$\sim20$ in spirals and flocculents, but only by a factor of $\sim2$ in
clump clusters. The decrease in clump mass for lower redshift is mostly
the result of a decrease in general galaxy brightness, which is a
cosmological selection effect. The distributions on the right suggest
the clump mass per unit galaxy light is nearly independent of
redshift.}\label{fig:gems_masses}\end{figure}

\begin{figure}\epsscale{.8}
\plotone{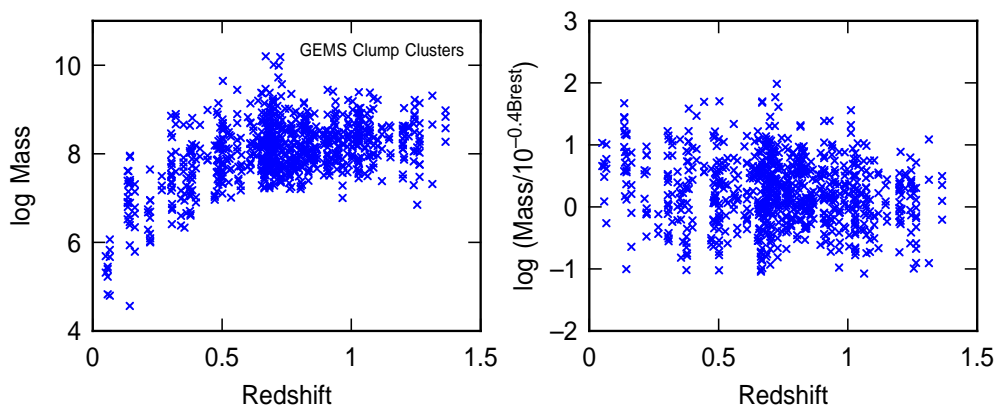} \caption{Clump mass (left) and ratio of clump mass to
total galaxy light (right) versus redshift for clumps in the GEMS
survey that could be measured in $V_{606}$ and $z_{850}$ passbands.
Background light from the interclump region is not subtracted from the
clump light in this case, so these masses are larger than the pure
star-formation masses plotted in the previous figure. The decrease in
clump mass for small redshift, combined with the constant clump mass
per unit galaxy light, indicate that the galaxies are suffering a
selection effect based on angular resolution and surface brightness,
but the clumps inside the galaxies are not.
}\label{fig:gems_masses_gems_galaxies}\end{figure}

\begin{figure}\epsscale{.65}
\plotone{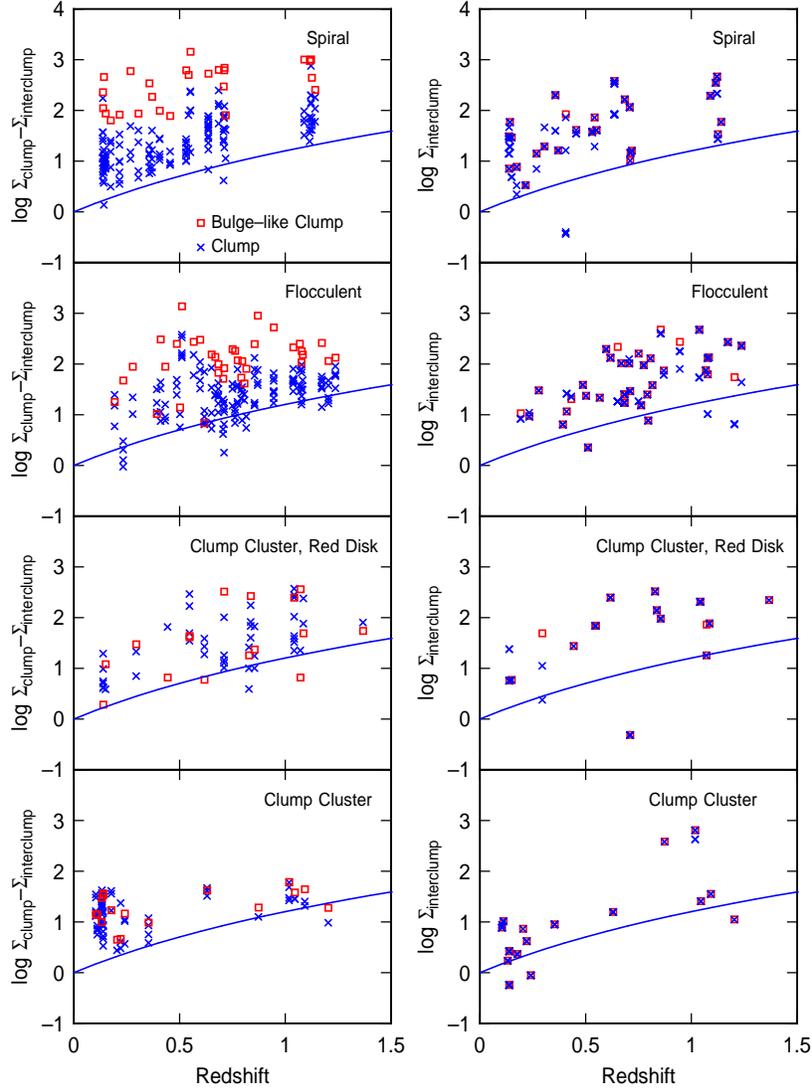} \caption{The redshift distribution of the excess
surface density of each clump (left) and the surface density of each
interclump region (right), both in $M_\odot$ pc$^{-2}$. Red squares are
for bulges or bulge-like clumps, blue crosses are for non-bulge clumps
(for an interclump region, this distinction is based on whether the
region was used to subtract the background for a bulge or a clump;
overlap of a cross and a square on the right means that an interclump
region was used for both). Bulges have higher surface densities than
clumps by a factor of $\sim7$ for spirals and flocculents, and about
the same surface densities as clumps for clump clusters. The curves
plot $\log(1+z)^4$, which is proportional to the log of the detection
limit. The bulges in spirals and flocculents have higher surface
densities than the bulges in clump clusters by a factor of $\sim6$. On
the right, the interclump surface density is a factor of $\sim3$ times
higher for spirals, flocculents, and clump clusters with red disks than
for clump clusters without red disks. This excess is consistent with
the conversion of clump clusters into spiral galaxies over
time.}\label{fig:gems_sb}\end{figure}

\begin{figure}\epsscale{1}
\plotone{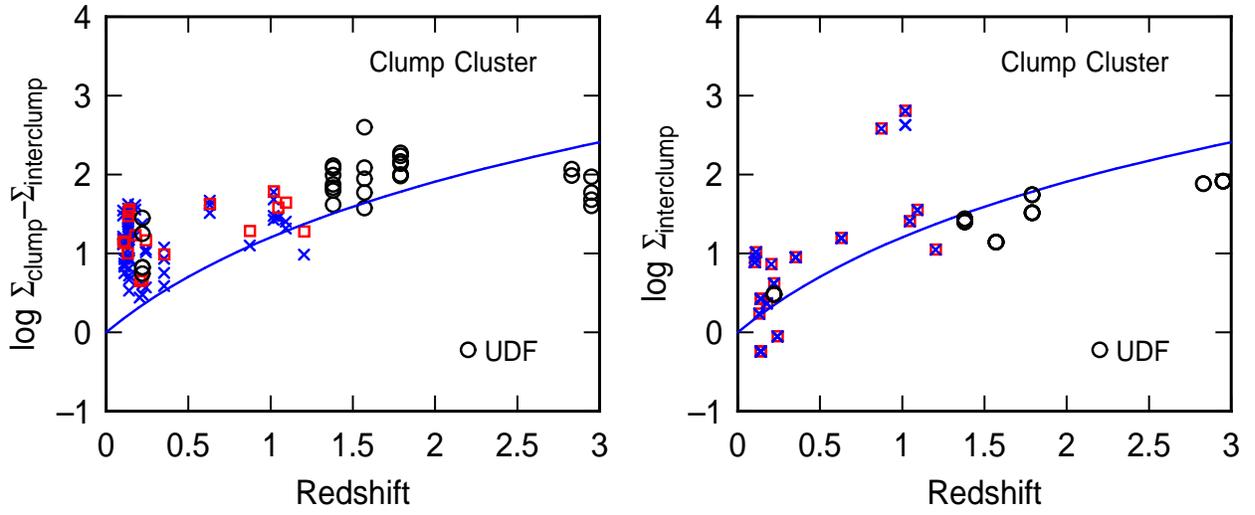} \caption{The redshift distribution of the excess
surface density of each clump (left) and the surface density of each
interclump region (right) for clump clusters with no obvious red disks,
combining the GOODS measurements from the previous figure with UDF
measurements from a small sample of clump clusters (open circles). The
curve traces a constant surface brightness limit.
}\label{fig:gems_sb_withudf}\end{figure}

\begin{figure}\epsscale{.65}
\plotone{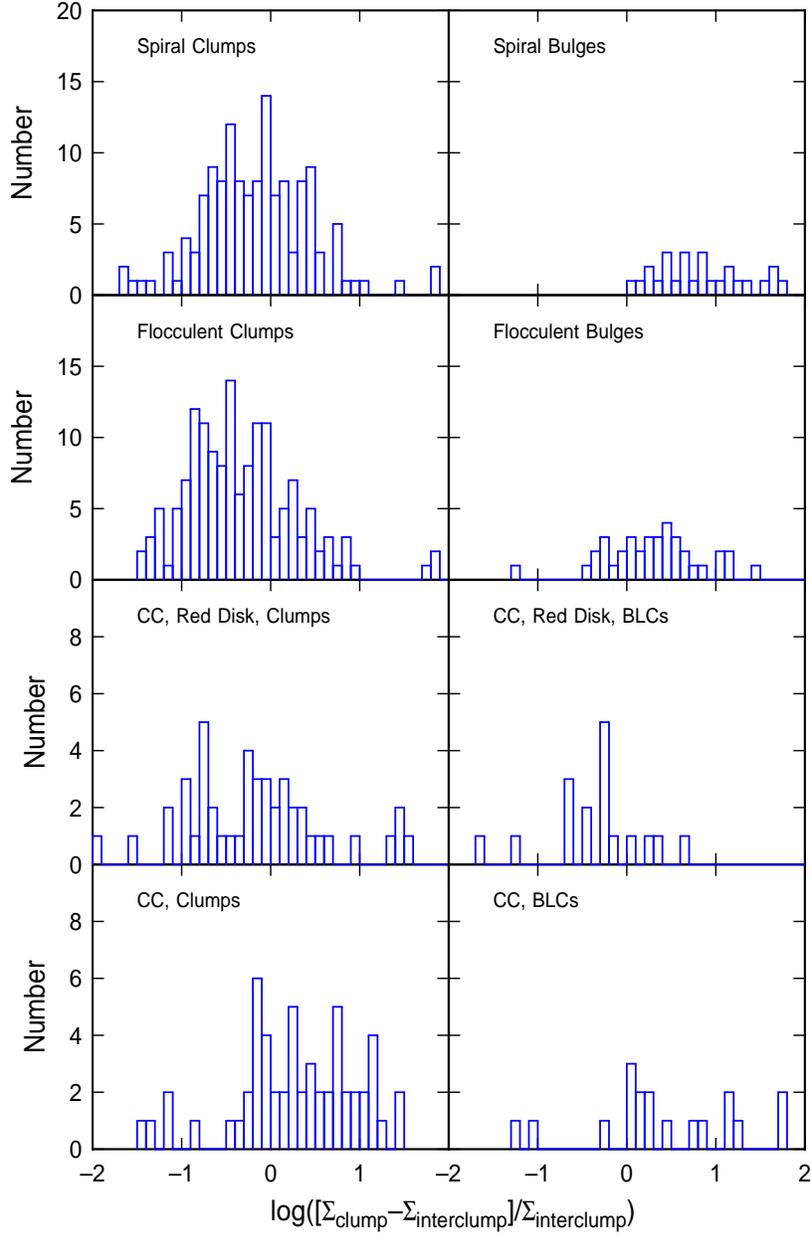} \caption{Histograms showing the ratio of the surface
mass density for the star-forming part of each clump to the surface
density of the interclump region, with non-bulge clumps on the left and
bulges or bulge-like clumps (BLC) on the right. The non-bulge clumps in
clump clusters have higher ratios than they do in spirals and
flocculents, which means that the clumps are more significant mass
perturbations in the clump clusters. Clump-cluster bulges have about
the same surface density contrasts as the clumps, while spiral and
flocculent bulges have much higher contrasts than the
clumps.}\label{fig:gems_ratio_histogram_comb}\end{figure}

\begin{figure}\epsscale{.8} \plotone{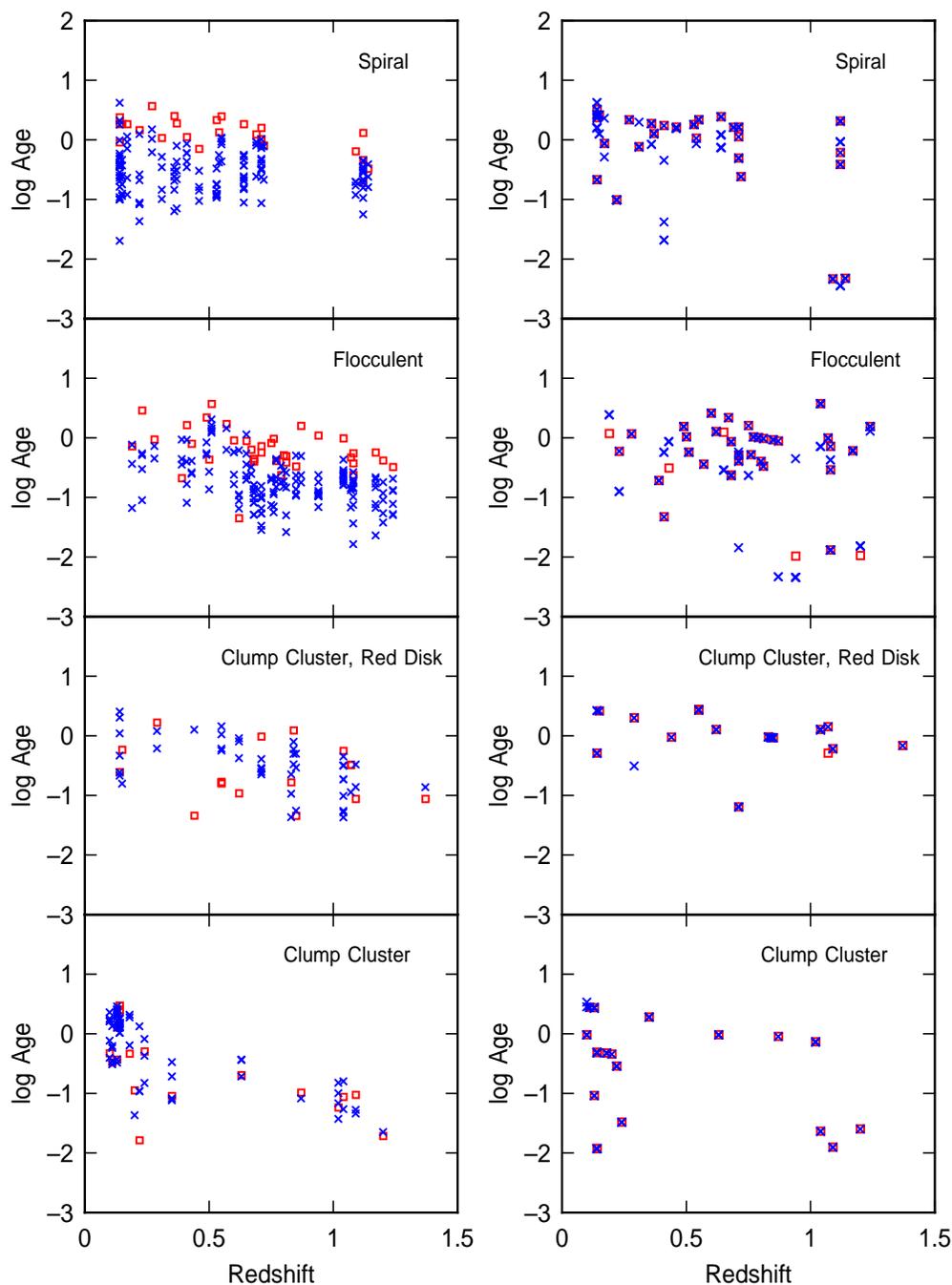} \caption{
Ages in Gyr of the excess emission from each clump (blue cross) and
bulge (red square) versus redshift (left) and ages of the interclump
regions (right).  Bulges are $\sim10\times$ older than clumps in
spirals and flocculents, but about the same age as clumps in clump
clusters.}\label{fig:gems_age}\end{figure}

\begin{figure}\epsscale{.7}
\plotone{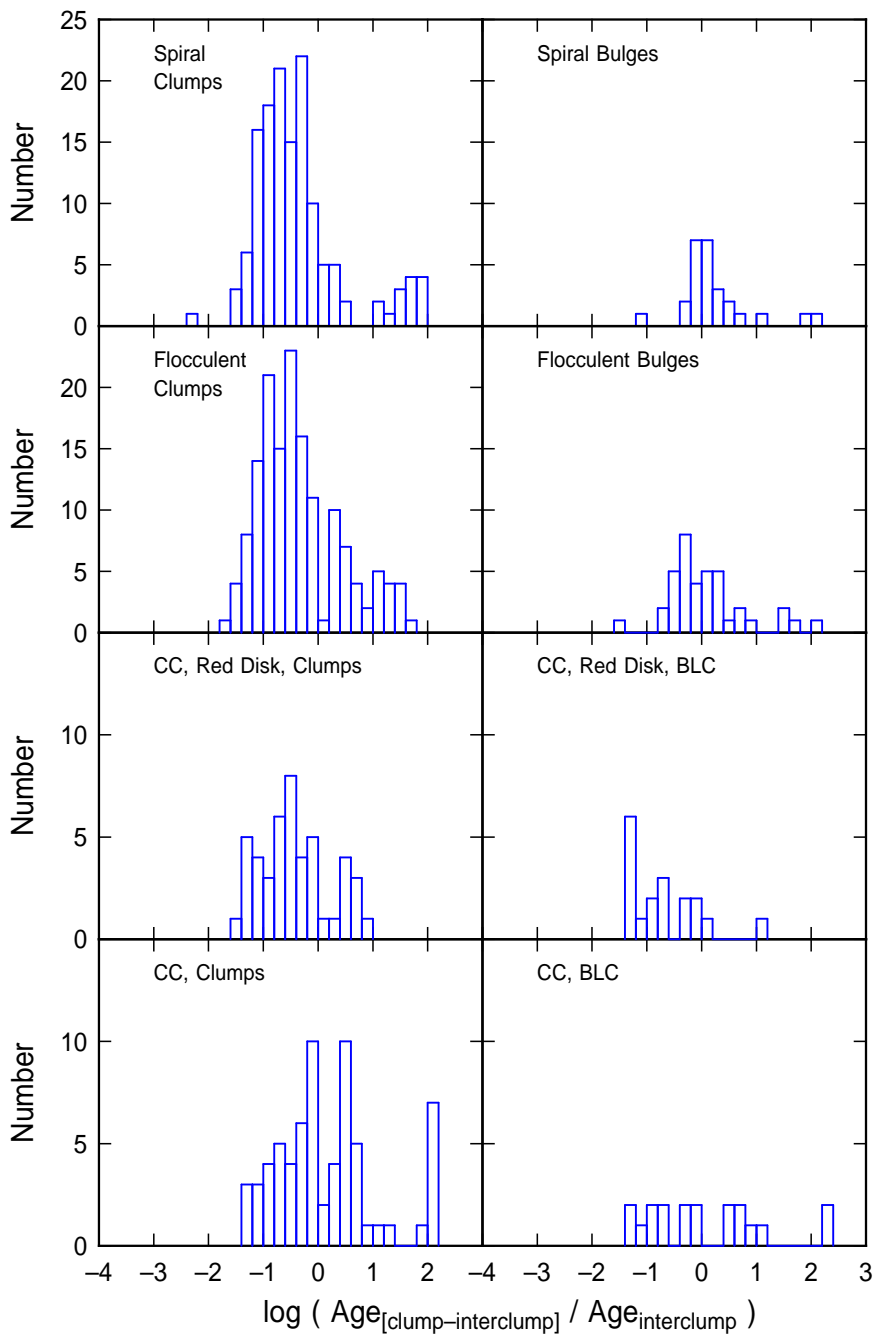} \caption{Histograms of the log of the ratio of the
age of the excess emission from each clump to the age of the associated
interclump region. Clumps are on the left and bulges or bulge-like
clumps (BLC) are on the right. In clump clusters without obvious red
disks, the clumps have about the same age as the interclump regions.
Clump clusters with red underlying disks and the spiral and flocculent
types have interclump regions older than the clumps by a factor of
$\sim3$. Bulges have about the same age as their nearby interclump
regions in all cases.}\label{fig:gems_age_ratio_his_comb}\end{figure}

\begin{figure}\epsscale{.7}
\plotone{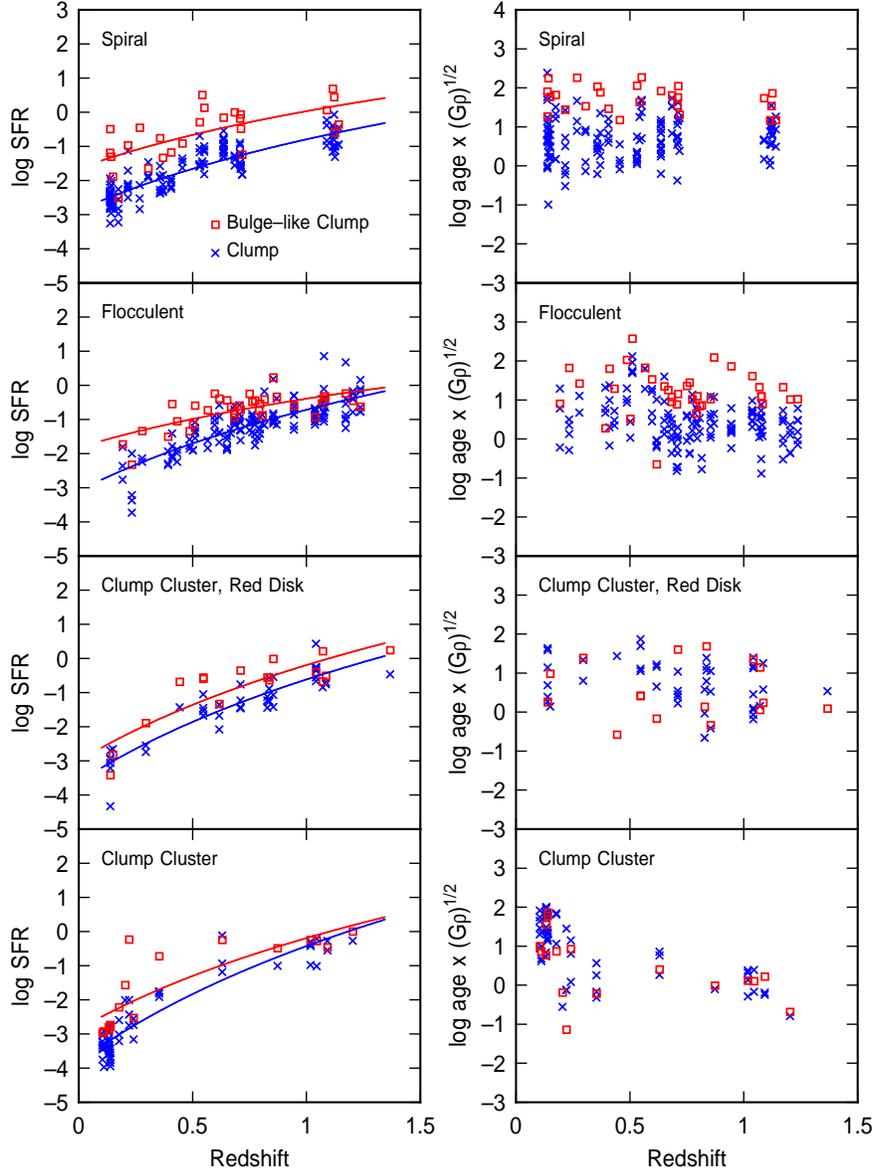} \caption{(Left) Clump star formation rates from the
ratio of mass above background to age (in $M_\odot\;{\rm yr}^{-1}$),
versus the redshift. The trends are fit with power laws that have an
average dependence of $SFR \propto\left(1+z\right)^{8}$; selection
effects are discussed in the text. (Right) The product of the clump age
and the clump dynamical rate is approximately constant over redshift
and averages about unity. Bulges are significantly older than clumps in
units of their dynamical time for spirals and
flocculents.}\label{fig:gems_sfr}\end{figure}

\begin{figure}\epsscale{.8}
\plotone{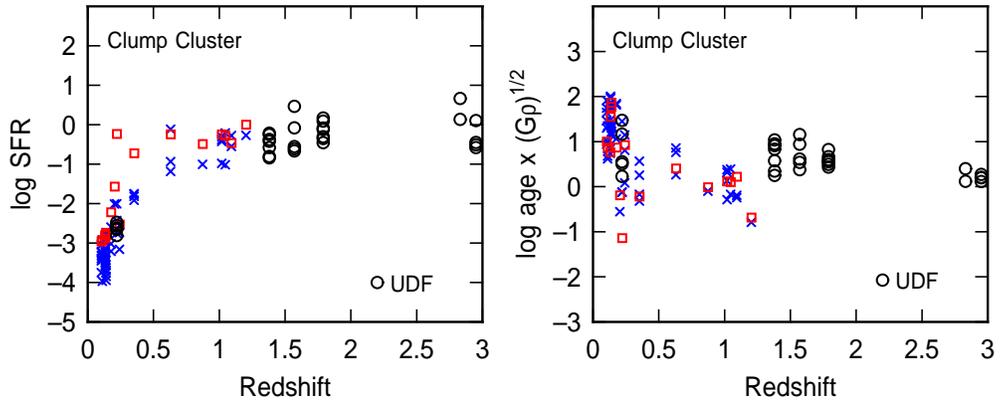} \caption{The average clump star formation rate for
clump clusters is extended to higher redshifts by including several UDF
galaxies measured and fit to models in the same
way.}\label{fig:gems_sfr_udf}\end{figure}

\begin{figure}\epsscale{.8}
\plotone{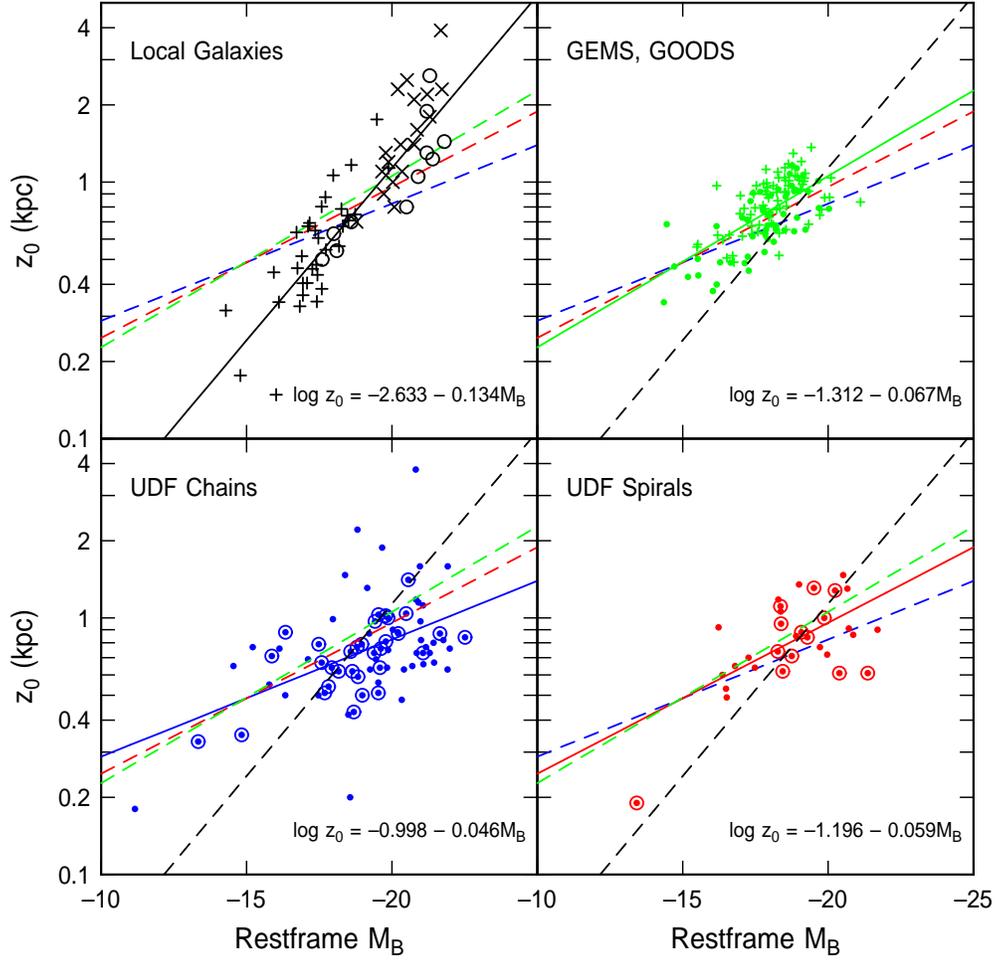} \caption{The scale height versus restframe absolute
magnitude is shown for local galaxies in the top left, GEMS and GOODS
spirals (plus symbols) and chains (dots) in the top right, UDF chains
in the bottom left and UDF spirals in the bottom right. For the UDF,
dots with circles are used for the best examples of these classes. Each
panel has a linear fit indicated by a solid line with the same color as
the points (and indicated by the equation), and it also has the fits
from the other panels in matching colors. For $M_B\sim-20$ mag, all
galaxies in these samples have about the same thickness, $\sim1$ kpc.
For GEMS, GOODS, and UDF, all measurements are in the $V_{606}$ band.
Local galaxies were measured in the R band by various
authors.}\label{fig:findHandMB}\end{figure}

\begin{figure}\epsscale{.8}
\plotone{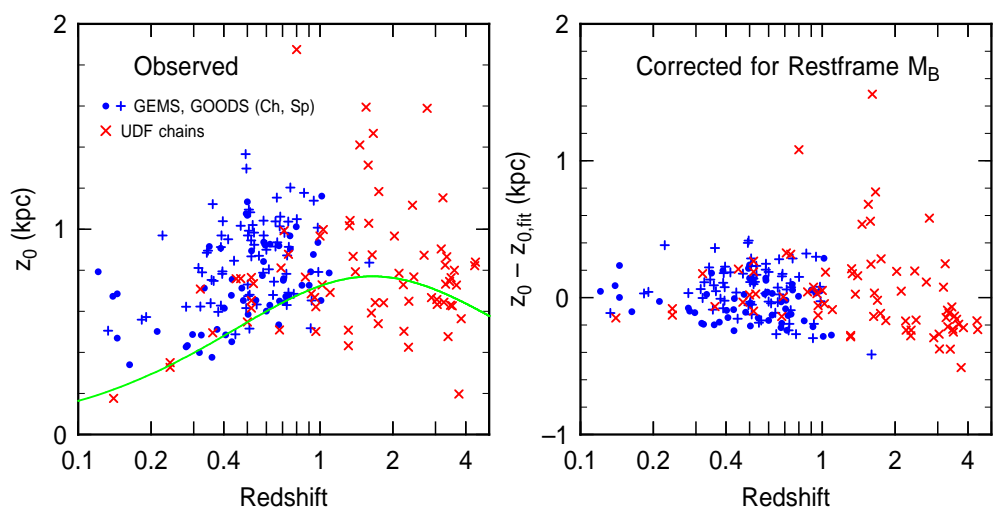} \caption{ (Left) Scale height versus redshift for
GEMS and GOODS spirals (plus symbols) and chain galaxies (dots) and UDF
chain galaxies (crosses). The decrease in $z_0$ for low redshift arises
because the galaxies are intrinsically fainter at lower redshifts. The
green curve shows the scale corresponding to 3 pixels, which is about
the FWHM of a point source in the GOODS image. (Right) The difference
between the measured scale height and the average scale height at the
same restframe $M_B$ is plotted versus redshift; there is no obvious
dependence.}\label{fig:findHandMB_vs_z_corrected_only}\end{figure}

There have been several studies of local galaxies blurred, dimmed, and
bandshifted to see what high redshift galaxies might look like (e.g.,
Brinchmann et al 1998; Burgarella et al. 2001; Smith et al. 2001;
Papovich et al. 2003; Taylor-Mager et al. 2007). Burgarella et al.
found that the largest change in morphology comes from viewing a galaxy
in the restframe UV, which makes a galaxy more asymmetric and less
centrally concentrated than in a visible band image.  In their examples
of redshifted local galaxies, however, the spiral arms are usually
still visible and the star-forming regions are only slightly higher
contrast than they are locally. They do not look like the clump
clusters shown here in Figures \ref{fig:fig_clump-red-no-060409} and
\ref{fig:Fig5-clumpy-060409}.

Overzier et al. (2008) showed that local compact UV-luminous galaxies
are similar to Lyman Break galaxies when convolved to the same spatial
resolution and viewed in the same restframe. They suggested that the
Lyman Break galaxies are collisional starbursts, like the local
galaxies.  It may be that some clump clusters are collisional
starbursts too. COMBO17 28751, 39638, 26313, 44885 and perhaps others
in Figure \ref{fig:Fig5-clumpy-060409}, have extended features that
could be tidal in origin. However, if chain galaxies are the edge-on
counterparts to clump clusters, then these galaxies are generally too
flat to be tidally distorted in a collision (Elmegreen \& Elmegreen
2006). We examined collisional galaxies in GOODS (Elmegreen et al.
2007b) and at these modest redshifts, they still look like local
collisions.  Also, the clump cluster UDF 6462 has extended features
like some clump clusters in Figure \ref{fig:Fig5-clumpy-060409}, but it
has a continuous rotation curve and a metallicity gradient, suggesting
it is a single clumpy disk (Bournaud et al. 2008).

\section{Discussion}
\label{sect:disc}

\subsection{Giant Clumps: Bandshifting, Selection Effects, and
Origins}\label{GC}

The GOODS field offers a view of young galaxy morphology at redshifts
$z<1$ with selection effects caused by bandshifting, variable spatial
resolution, and variable surface brightness dimming. The effects of
bandshifting are not so bad at these redshifts, though. GOODS galaxies
observed in the $z_{850}$ ACS band are bandshifted only to their
restframe V or B-band, where we know what local galaxies look like. UDF
bandshifting for $z\sim2-3$ galaxies is much worse, as it takes a
$z_{850}$ image into the restframe UV, where even the local
morphologies are uncertain and, in some cases, quite different than in
the optical bands. Variable spatial resolution is more of a problem for
GOODS than the UDF because the spatial scale per pixel increases
strongly with redshift at $z<1.6$; the increase slows and reverses
beyond that. Surface brightness dimming is a problem in both near and
far redshift surveys, limiting what can be seen to the brightest
resolved features and producing a strong correlation between measured
surface brightness and redshift. Here we discuss several properties of
young disk galaxies that are relatively insensitive to these three
selection effects.

First of all, the two new morphologies found at $z\sim2$ in the UDF,
i.e., chain galaxies and clump clusters, are still present in GOODS in
the redshift range from 0 to 1, alongside normal-looking spirals. This
demonstrates that bandshifting alone does not cause the appearance of
clumpy structure. Clumpy galaxies still have no spirals or regular
exponential disks in GOODS, and about half still have no bulges. The
clump/interclump contrast in total mass surface density is large for
these galaxies, e.g., $\sim2-5$, even in the restframe B-band. It is
significantly smaller, $\sim1.1-2$, for spirals and flocculent galaxies
at the same redshift in GOODS. Bulges are more massive than
star-forming clumps in spirals by a factor of $\sim16$ and older by a
factor of $\sim10$, whereas bulges are more massive than clumps in
clump clusters by a factor of only 2.2 and they are not significantly
older. Spirals and flocculents have higher interclump mass surface
densities than clump clusters too, by a factor of $\sim3$ at the same
redshift. All of these results suggest that clumpy galaxies are younger
versions of spiral and flocculent galaxies, and that this youthful
appearance extends even to objects observed at recent cosmological
times ($z<0.2$).

The redshift dependencies of surface brightness, physical resolution,
and restframe wavelength all contribute to a strong redshift dependence
for the choice of galaxy in this survey, and ultimately, to the derived
average star formation rate in a clump, given that clump mass scales
with galaxy luminosity. This rate therefore has little utility in
understanding the star formation process. However, our result that the
age of a clump is comparable to its dynamical time, as determined by
the average clump density, is relatively insensitive to these selection
effects.  This result confirms our assumption that the clumps are
star-forming regions and it suggests that star formation is a dynamical
process involving disk self-gravity, as it is locally.

The similarity between the star formation age and the dynamical time
also makes a strong statement about the origin of the clumps: they are
not separate galaxies brought in from outside and settling into a
common disk.  If they were, then their background-subtracted ages would
be significantly larger than their dynamical times, i.e., they would be
older, self-bound, and more independent of each other.

The suggested origin of disk clumps by gravitational instabilities is
analogous to the process commonly thought to trigger large-scale star
formation in local galaxies (e.g., Elmegreen 2002). If we take the
analogy to local galaxies further, then we can compare the largest
scales of star formation in the two cases. For local galaxies, the
largest scale of coherent star formation is about the disk Jeans
length, $L_J=\sigma^2/\left(\pi G\Sigma\right)$, for velocity
dispersion $\sigma$ and mass column density $\Sigma$ (which includes
stars if the stellar and gaseous dispersions are comparable). The mass
on this scale is about the Jeans mass,
$M_J=\sigma^4/\left(G^2\Sigma\right)$. That this is a characteristic
scale for star formation and not a preferentially sampled scale has
been shown by power spectra of optical galaxy images (Elmegreen et al.
2003a,b), fractal structure analysis of optical and HII region images
(Elmegreen et al. 2006), and autocorrelation analysis of cluster
positions (Zhang et al. 2001). In these studies, the structure involved
with star formation itself is scale free, i.e., it has a power law
power spectrum, but that power law only extends up to a scale of about
1 kpc, which is the characteristic or outer scale. Beyond that,
star-forming regions appear somewhat independent and uncorrelated. We
also note that for star formation on a dynamical time, the largest
region that has a clump-like shape rather than a spiral shape,
distorted by shear, is this same Jeans length (Elmegreen \& Efremov
1996). This is because the dynamical time is less than the shear time
on scales smaller than $L_J$ in a marginally stable disk. Thus, whether
a star-forming clump is the result of interstellar condensation from a
gravitational instability or the result of widespread turbulence
compression in a shearing environment, there is an outer scale
comparable to $L_J$.

In the context of high redshift galaxies, there is a selection effect
where we only see structures larger than the limiting spatial
resolution. This structure would not generally have a characteristic
length representative of the star-formation process, and in fact we
cannot measure such a length if the structure is unresolved. However,
we have the fortunate circumstance for young galaxies that the contrast
between the bright features (clumps) and the regions between the bright
features (interclump stars) is very large, giving these galaxies the
appearance, even in optical restframes (Fig.
\ref{fig:fig_clump-red-no-060409} and \ref{fig:Fig5-clumpy-060409}), of
extreme clumpiness. Then the clump masses can be measured from their
luminosities and colors even if the clumps are unresolved, without
severe blending problems.  To demonstrate this difference with local
galaxies, we compared in Figure \ref{fig:fig19-N7793,GOODS-060409} a
local flocculent galaxy with two clumpy GOODS galaxies, viewed with the
same spatial resolution and restframe wavelength.  In the local galaxy,
star formation patches on the scale of $L_J$ are everywhere in an
exponential disk, and their contrast is not particularly large. In the
clumpy GOODS galaxy, however, there are only a few very bright regions
that are well separated from each other. In this sense, they are
resolvable (from each other) and measurable in mass. After subtracting
the light from the surrounding disk, we found that the star-forming
parts of these clumps have masses of $10^7$ to $10^8\;M_\odot$ in young
stars, with a few clumps as massive as $10^9\;M_\odot$.  If there is
gas in these clumps as well as stars, then the masses would be larger,
perhaps by a factor of 2 or more. Because we identify this outer-scale
mass with $M_J$, by analogy with local galaxies, we conclude that $M_J$
is larger for star-forming clumps in the clumpiest high redshift
galaxies than it is in spiral galaxies locally. This case is more
compelling for the GOODS clump clusters than the UDF clump clusters
(EEFL) because bandshifting is not as severe for the GOODS galaxies,
which means that the morphological contrast to spiral galaxies is more
clear.

There is still a surface brightness limit in the GOODS and UDF clump
clusters that limits the clumps we can measure to only those with the
highest surface densities. There should be many more fainter and
smaller young regions in clump cluster galaxies than we can observe in
these surveys. Thus we know little about typical star-forming regions
or luminosity functions. Our conclusion is only that the maximum mass
of a coherent unit of star formation increases with redshift. If we
identify this mass with $M_J$, as proposed above, then such a result
would follow most sensitively from an increase in $\sigma$, the
velocity dispersion of the ambient neutral medium.  We have made the
same point for UDF galaxies before (EEFL) and noted how observations of
random gas motions tend to support this higher dispersion
(F{\"o}rster-Schreiber et al. 2006; Weiner et al. 2006; Genzel et al.
2006, 2008), although the dispersion in the ambient neutral medium has
not yet been measured with high angular resolution.

\subsection{On Clump Migration to make a Bulge}

We are interested in whether the clumps we observe are so massive and
dense compared to their surrounding disks that they interact with each
other and the halo, losing angular momentum and spiraling into the
center (e.g., Noguchi 1999; Immeli et al. 2004a,b; BEE). To study this,
we estimate the ratio of clump mass to galaxy mass and compare this
with the ratio in simulations where the clumps do migrate to the
center.  In BEE, we found that the total clump mass was $\sim30$\% of
the disk mass (gas+stars), and the simulations formed $\sim6$ giant
clumps which moved to the center in $\sim1$ Gyr. Thus each clump was
$\sim5$\% of the disk mass.  For the GOODS galaxies, we use the
right-hand side of Figure \ref{fig:gems_masses}, which shows the ratio
of the clump mass to a measure of the galaxy luminosity,
$10^{-0.4Brest}$.  To convert this luminosity to mass, we again use the
population evolution models in Bruzual \& Charlot (2003) for a Chabrier
IMF and a metallicity of 0.4 solar.  Recall from Section
\ref{thickness} that $M_B=4.88+2.37\log(T)$ mag per unit solar mass of
stars with age $T$ in Gyrs.  The residual stellar mass of this
population varies with age as $M=0.60\times10^{-0.078\log T}M_\odot$.
If we write $M=A10^{-0.4M_B}$ in analogy with the formulation in Figure
\ref{fig:gems_masses}, then we derive $A=50\times10^{0.87\log
T}\;M_\odot$. As a consistency check, note that this gives a galaxy
luminous mass $M=2.7\times10^{10}\;M_\odot$ for $Brest=-20.3$ mag and a
mean population age of $T=5$ Gyr. The total galaxy mass would be larger
because of dark matter.

In Figure \ref{fig:gems_masses},
$M_c\sim10^{-0.1\pm0.4}\times10^{-0.4Brest}\;M_\odot$ for clump masses
$M_c$ in clump clusters with no obvious underlying disk.  The ratio of
the clump mass to the galaxy mass is therefore
$10^{-0.1\pm0.4}/A=10^{-1.8\pm0.4 - 0.87\log T}$ for average galaxy
population age $T$, in Gyr.  For the average clump studied here, this
mass ratio ($=1.6$\%) is smaller than the ratio for the BEE simulations
(5\%) by a factor of $\sim3$ for $T=1$. Clumps that are more massive
than average by one standard deviation have a clump-to-galaxy mass
ratio of 4\%, which is close to the simulation ratio. Considering that
the BEE simulations did not determine a mass limit for accretion to the
center but only had clump masses that automatically came from the disk
instability, the observed clumps in clump cluster galaxies could be
massive enough to spiral in for some of the distance, maybe even to the
center if they are also dense enough to withstand the higher tidal
forces there.  The clumps in the spiral and flocculent galaxies studied
here are considerably lower in mass than this value for clump clusters,
by another factor of 6, and are therefore not likely to spiral in
significantly. The spiral clumps will probably disperse where they are,
as in modern spiral and flocculent galaxies.

\subsection{Clump Clusters as Massive versions of Local Dwarf
Irregulars} \label{dwarf}

The comparisons in section \ref{sect:7793} between two local galaxies
without prominent spiral density waves and GOODS clump clusters, viewed
at the same spatial resolution and rest wavelength, illustrate two key
differences. First, for similar size galaxies (flocculents versus clump
clusters), the GOODS galaxies have higher intrinsic clump/interclump
brightness contrasts. Second, for similar morphologies (dwarf
Irregulars versus clump clusters), the GOODS galaxies are intrinsically
brighter by a factor of 10 to 100.

Evidently, some clump clusters resemble massive versions of local dwarf
Irregulars. If the star-forming regions in both result from
gravitational instabilities, then the clumps have to be relatively
large in each type. This is presumably because of a relatively high gas
fraction and a relatively large Jeans length compared to the galaxy
size. The Jeans length in a gas-rich system is $L_J=\sigma^2/\pi G
\Sigma$ for turbulent speed $\sigma$ and mass column density $\Sigma$.
For a disk-to-halo mass ratio $f_D$, $\pi
G\Sigma=V^2/\left(R\left[1+1/f_D\right]\right)$ for rotation speed $V$
and disk scale $R$. Thus $L_J/R\sim\left(\sigma/V\right)^2$ for
moderate to large $f_D$. When $L_J/R$ is large, $\sigma/V$ must be
large, and there is a small number of relatively large star formation
clumps in a galaxy.

A large gravitational length should also be evident in the disk
thickness, which is the same as $L_J=\sigma^2/\pi G\Sigma$ for a
single-component system, or $\sim\sigma^2/\pi G \Sigma_{total}$ for the
gas in a gas+star system with total $\Sigma_{total}$ inside the gas
layer. For dwarf Irregulars, both the clump size and the disk thickness
are indeed large compared to the radial scale length (Binggeli \&
Popescu 1995; Sung et al. 1998). The thicknesses of clump clusters are
difficult to determine because of selection and resolution effects, but
they appear to be large compared to the radial sizes too (Sect.
\ref{thickness}).

Dwarf Irregulars have both a thick disk and a clumpy structure because
of their large value of $\sigma/V$, which results from a small rotation
speed $V<100$ km s$^{-1}$ and a normal dispersion, $\sigma\sim10$ km
s$^{-1}$. Clump clusters are apparently massive galaxies with about the
same ratio of $L_J/R$ and $\sigma/V$, but in this case, $V$ is probably
typical of $M^*$ galaxies, namely $V\sim150-200$ km s$^{-1}$. This
means that the gaseous velocity dispersion has to be high, $>20$ km
s$^{-1}$. This is not unrealistic considering observations of high
velocity dispersions in the ionized gas (Sect. \ref{intro}) and other
constraints mentioned in the previous two subsections.

Another similarity between clump clusters and dwarf Irregulars is that
neither have prominent spiral arms. Locally, this characteristic
results from a lack of strong tidal or other asymmetric forces and, in
the case of dwarfs, from a large stellar velocity dispersion compared
to the rotation speed. The first of these points suggests that when
there are no tidal arms the clump cluster morphology results from
internal processes.  This emphasizes again a likely analogy to local
dwarf Irregulars, rather than local mergers.

Collisions are necessary in the local universe to make a superstarburst
because collisions are the only way that a local galaxy can rapidly
accrete a large amount of gas. The primary ingredient for a
superstarburst is rapid gas accretion -- much faster than the gas
consumption rate by star formation, which is only several percent of
the galaxy-wide dynamical rate. Interactions can bring in gas to the
inner disk of a galaxy through tidal torques and through direct
contact, perhaps doubling the gas mass surface density in one dynamical
time. This is much faster than the burn-off rate from normal star
formation. At high redshift, however, cosmological accretion through a
cold flow might double the amount of gas in a dynamical time without
any other galaxy involved (Sect. \ref{intro}). The resulting gas and
starburst will generally have an irregular structure as the cold flow
is not likely to be symmetric and the young disk is likely to be
unstable. Thus it is possible that a high fraction of clump cluster
galaxies, and perhaps some clumpy Lyman Break galaxies too, have their
morphology and large star formation rates because of a high gas
accretion rate and a high gas turbulent speed in an intrinsically dense
disk, rather than because of a current merger. This makes them
high-density and high-mass analogs of local dwarf Irregulars, rather
than high-redshift analogs of ULIRGS.

Clumpy galaxies are observed in a young state, so they can have a high
gas fraction and high state of turbulence. The rate of star formation
is high in them because the density is high, unlike the situation in
local dwarf Irregulars where the intrinsic density is low and the star
formation rate is low.

\subsection{Disk Stability and Damped Lyman Alpha Limits on the Star Formation Rate}

A high velocity dispersion normally stabilizes a disk but clump
clusters also need a high gas column density to simultaneously match
the sizes and the masses of the giant clumps (EEFL). Recall that the
Jeans length scales with $\sigma^2/\Sigma$ and the Jeans mass scales
with $\sigma^4/\Sigma$, so the mass per unit length scales with
$\sigma^2$. Regions that are larger by a factor of $\sim3$ and more
massive by a factor of $\sim100$ require a dispersion that is larger by
a factor of $\sim5$ and a mass column density that is larger by a
factor of $\sim10$ (EEFL).  They are unstable even with the high
dispersion. The required $\Sigma\sim100\;M_\odot$ pc$^{-2}$ is
comparable to the column density in the inner parts of modern spirals,
and suggests again that clump clusters are young, gas-rich versions of
local galaxies.

The high dispersion suggests a solution to the problem raised by Wolfe
\& Chen (2006), that damped Lyman alpha absorption (DLA) in quasars
often indicates a column density of HI ($N>2\times10^{20}$ cm$^{-2}$)
that is unstable in the local universe but has no associated star
formation in the high redshift universe. Wolfe \& Chen noted that the
observed gas has to be at least 10 times less efficient at forming
stars than local galaxies, according to the Kennicutt (1998) relation.
The local star formation threshold is $\sim5\;M_\odot$ pc$^{-2}$
($\sim2.6\times10^{20}$ H cm$^{-2}$ including He). Many DLA systems
have higher column densities than this with no evident emission.

Wolfe \& Chen suggested that the column density threshold could be high
in DLA galaxies because at high redshift only the dense inner parts of
galaxies are well formed, and these parts have high angular rotation
rates (i.e., for a fixed galaxy density relative to the average density
of the universe, the angular rotation rate scales approximately
inversely with the Universe's age). High angular rotation rates
stabilize the gas through the epicyclic frequency $\kappa$ in the
Toomre expression for the critical column density,
$\Sigma_{crit}=\sigma\kappa/\left(3.36G\right)$.  In this
interpretation, stability occurs essentially because the galaxies are
small.  Our observations of clump clusters in the UDF do not find a
size that correlates well with redshift (Elmegreen et al. 2007a), but
we agree in principle with the Wolfe \& Chen suggestion that the most
active parts of clump clusters probably correspond to the inner regions
of today's disks, primarily because that is all we can observe at the
surface brightness limit.

Wolfe \& Chen also suggested that the molecular fraction could be low
in DLA gas as a result of low metallicities, thereby requiring higher
$\Sigma_{crit}$ to get molecules and star formation. Local dwarf
irregulars have low metallicities and molecular fractions too, but
$\Sigma_{crit}$ is lower for them than it is for spirals (Hunter et al.
1998). Other stabilizing mechanisms such as disk flaring or a lack of
cold gas were ruled out by Wolfe \& Chen.

The present observations suggest an additional solution to this
problem. Young galaxies seem to have higher turbulent speeds than
modern galaxies by a factor of $\sim5$ or more, and $\Sigma_{crit}$
increases in direct proportion to this speed.  When star formation
occurs in the unstable part of a gas-rich, highly-turbulent disk, it
should be fast, violent, and make massive star complexes.  This is what
we observe in clump clusters. For the same dispersion, regions with
lower column densities should be more stable and relatively quiescent.
This is apparently what Wolfe \& Chen find. We suggested in Section
\ref{dwarf} that the turbulent speed in the disks of young massive
galaxies is $\sim20$ km s$^{-1}$ or more. This makes them stable at
column densities that would be unstable in local spirals. Measurements
of DLA line widths (Wolfe \& Prochaska 1998) include cases with such
high values, but the overall DLA profile could be contaminated by disk
rotation, making the turbulent speed uncertain.

If we consider star-forming instabilities in the context of clump
cluster morphology, we can use the analogy with local dwarfs to infer
that $\sigma/V\sim\left(L_J/R\right)^{1/2}$ for velocity dispersion
$\sigma$, rotation speed $V$, star formation scale $L_J$, and galaxy
size $R$ (Sect. \ref{dwarf}). Then the critical column density is
$\Sigma_{crit}\sim \left(L_J/R\right)^{1/2} \left(V^2/R\right)/1.7G$
for $\kappa\sim 2V/R$ in the case of solid body rotation (use
$\kappa=1.4V/R$ for a flat rotation curve). With equally clumpy
morphologies, $\left(L_J/R\right)^{1/2}$ should be the same for clump
clusters and local dwarf irregulars, making $\Sigma_{crit}$ scale with
the square of the rotation speed. It should therefore be much larger in
turbulent, high-redshift galaxies of normal size than it is in local
dwarfs. In a galaxy with less clumpy structure, such as local spirals,
$\Sigma_{crit}$ should be smaller for the same $V$ because
$\left(L_J/R\right)^{1/2}$ is small. Local dwarfs have the lowest
$\Sigma_{crit}$ because their lower $V$ offsets the increase in
$\left(L_J/R\right)^{1/2}$.

\section{Conclusions}

Clumpy galaxies have been examined in GOODS and GEMS and their clump
properties and disk thicknesses measured. We are interested in the
transition between these irregular types and modern disk systems. Our
results may be summarized as follows:

1. Chains and clump clusters are present at photometric redshifts down
to $\sim0.1$ or lower, along with spiral galaxies with the same
magnitudes and redshifts.  This observation indicates that the clumpy
morphology is not the result of bandshifting. That is, chains and clump
clusters are not normal spiral galaxies simply viewed in the extreme
ultraviolet restframe.  There is a tendency for star formation to look
more clumpy at shorter wavelengths, but the clump cluster morphology is
generally more extreme than that.

2. The primary difference between clumpy galaxies and local spiral
galaxies is the contrast in both intensity and mass surface density
between the clumps and the interclump regions. The clump clusters
studied here have contrasts in mass surface density between the young
parts of the brightest clumps and the surrounding interclump regions
that are factors of 1 to 4, which means that the total contrasts in
mass surface density are factors of 2 to 5. Spiral and flocculent
galaxies at the same redshifts as the clump clusters have much smaller
clump contrasts, 0.1-1 for the young parts of the clumps, or 1.1-2 for
the total.

3. There appears to be an evolutionary sequence from clump clusters
with no evident red underlying disks, to clump clusters with red
underlying disks and in some cases bulges, to spiral galaxies with
either flocculent or long-arm spiral structures. Along this sequence,
the clump/interclump surface density contrast decreases, and bulges
appear with greater distinction from the clumps in terms of mass,
surface density, and age. There are no evident external processes or
merger-like processes, tidal tails, etc., associated with this change
in bulge and clump morphology, suggesting that bulges grow from
internal disk processes. Such processes might include clump coalescence
and loss of clump angular momentum, as driven by gravitational friction
and asymmetrical forces (BEE).

4. This evolutionary sequence is mixed in redshift for the GOODS
sample, which means that the morphologically youngest galaxies, the
clump clusters with no evident interclump emission, have about the same
redshift distribution as the morphologically oldest galaxies, the
spirals. This mixture implies that clump clusters are either
intrinsically young, and therefore form continuously from intergalactic
gas over a wide range of redshifts, or clump clusters rejuvenate from
faint unseen forms to the starbursting clumpy systems that we see,
possibly following a major gas accretion event. Considering the
resemblance between clump clusters and dwarf Irregulars discussed in
this paper, the faint unseen forms could be massive analogs of local,
low surface brightness, dwarf Irregulars.

5. There is evidence for tidal structures in some clumpy galaxies, but
not in all. Generally, the clump cluster morphology is distinct from
the morphology of interacting galaxies.  Many other galaxies in GOODS
and GEMS covering the same redshift range as clump clusters are clearly
interacting, showing all the usual signs of interactions, such as tidal
tails, tidal debris, and rings (Elmegreen et al. 2007b). Thus clump
clusters are not merger remnants whose tidal debris has been suppressed
by cosmological surface brightness dimming. They are either somewhat
isolated, or they are interacting less frequently and less strongly
than conventional mergers. This, along with certain morphological
details in a clumpy galaxy's structure, suggest that galaxy growth is
dominated by smooth gaseous inflow and not the merger of smaller
galaxies (Bournaud \& Elmegreen 2009). Recent numerical simulations of
galaxy formation in a cosmological context reinforce this
interpretation (Sect. \ref{intro}). Clumpy asymmetric structure in a
high redshift galaxy does not necessarily imply a merger.

6. Clump clusters resemble local dwarf Irregulars in restframe
morphology far better than they resemble flocculent spirals and
mergers. However, clump clusters have the luminosities and masses of
normal spiral galaxies, which are 10 to 100 times larger than local
dwarf Irregulars.  Thus clump clusters represent a unique galaxy stage:
they are as massive as normal spirals but as irregular as dwarfs. This
unique property is probably connected with their extreme youth.
Although neutral gas in clump clusters is not widely observed yet, we
feel confident in predicting that these galaxies will be found to have
high gas fractions, as do dwarf Irregulars, and high gas velocity
dispersions relative to their rotation speeds, as do dwarf Irregulars.
The velocity dispersions of the ionized gas components are already
observed to be high (Sect. \ref{intro}).

7. Local dwarf Irregulars are not a perfect analog to high redshift
clumpy galaxies.  The dwarfs have low star formation rates and evolve
slowly, whereas the high redshift systems are massive, star-bursting,
and in extreme cases, evolve quickly to symmetric galaxies with bulges
and exponential disks (BEE). The slow evolution for local dwarfs
follows in part from their low disk surface density, and this helps
explain why they still have high gas fractions after a Hubble time. The
difference between the two cases seems to be partly a matter of scale
-- not an indication of different physical processes. Scaling issues
have selection effects that depend the epoch of the system and the
sensitivity of the observations. At early times, massive disk galaxies
take the form of clump clusters and can be as unevolved as today's
dwarf Irregulars. Lower mass galaxies at these early times would not be
observable. At late times, the massive clumpy disks have evolved into
smooth disks, and the low mass versions, the dwarf Irregulars, are the
only visible remnants of this phase.

8. The masses of star-forming regions relative to the surrounding
galaxy appear to be larger by a factor of $\sim6$ in clump clusters
than in spirals at the same redshift (Sect. \ref{masses}). The clump
masses also increase with redshift, although selection and resolution
effects could contribute somewhat to this mass increase. More likely,
the increase in relative clump mass parallels the observed increase in
relative clump separation that defines the clumpy morphology (i.e.,
compared to the galaxy radius), and both are the result of an increase
in gas turbulent speed relative to the galaxy rotation speed.  The
turbulent speed of the neutral gas component in high redshift galaxies
is not yet observed, but it is predicted to be high, $20-50$ km
s$^{-1}$.

9. A high gas velocity dispersion relatively to the rotation speed
increases the threshold column density for gravitational instabilities.
This can explain the observed lack of star formation in gas that
produces damped Lyman alpha absorption.  A comparable ratio of
dispersion to rotation speed in local dwarf Irregulars and in high
redshift clump clusters, as suggested by their similar morphologies,
also explains the low threshold column density for star formation in
local dwarfs.

We are grateful to the referee for useful comments. DME thanks the
National Science Foundation for research support for KS and MP through
the Keck Northeast Astronomy Consortium REU program, and Vassar College
for support for MM and AY through the Undergraduate Research Summer
Institute (URSI) program, and for publication support through the
Research Committee and the Maud Makemson fund. Some of the data
presented in this paper were obtained from the Multimission Archive at
the Space Telescope Science Institute (MAST). STScI is operated by the
Association of Universities for Research in Astronomy, Inc., under NASA
contract NAS5-26555. Support for MAST for non-HST data is provided by
the NASA Office of Space Science via grant NAG5-7584 and by other
grants and contracts. This research has made use of the NASA/IPAC
Extragalactic Database (NED) which is operated by the Jet Propulsion
Laboratory, California Institute of Technology, under contract with the
National Aeronautics and Space Administration.

\clearpage

\clearpage

\clearpage

\clearpage

\clearpage

\clearpage

\clearpage

\clearpage

\clearpage

\clearpage

\clearpage

\clearpage

\clearpage

\clearpage

\clearpage

\clearpage

\clearpage

\clearpage

\clearpage
\begin{figure}\epsscale{.9}
\plotone{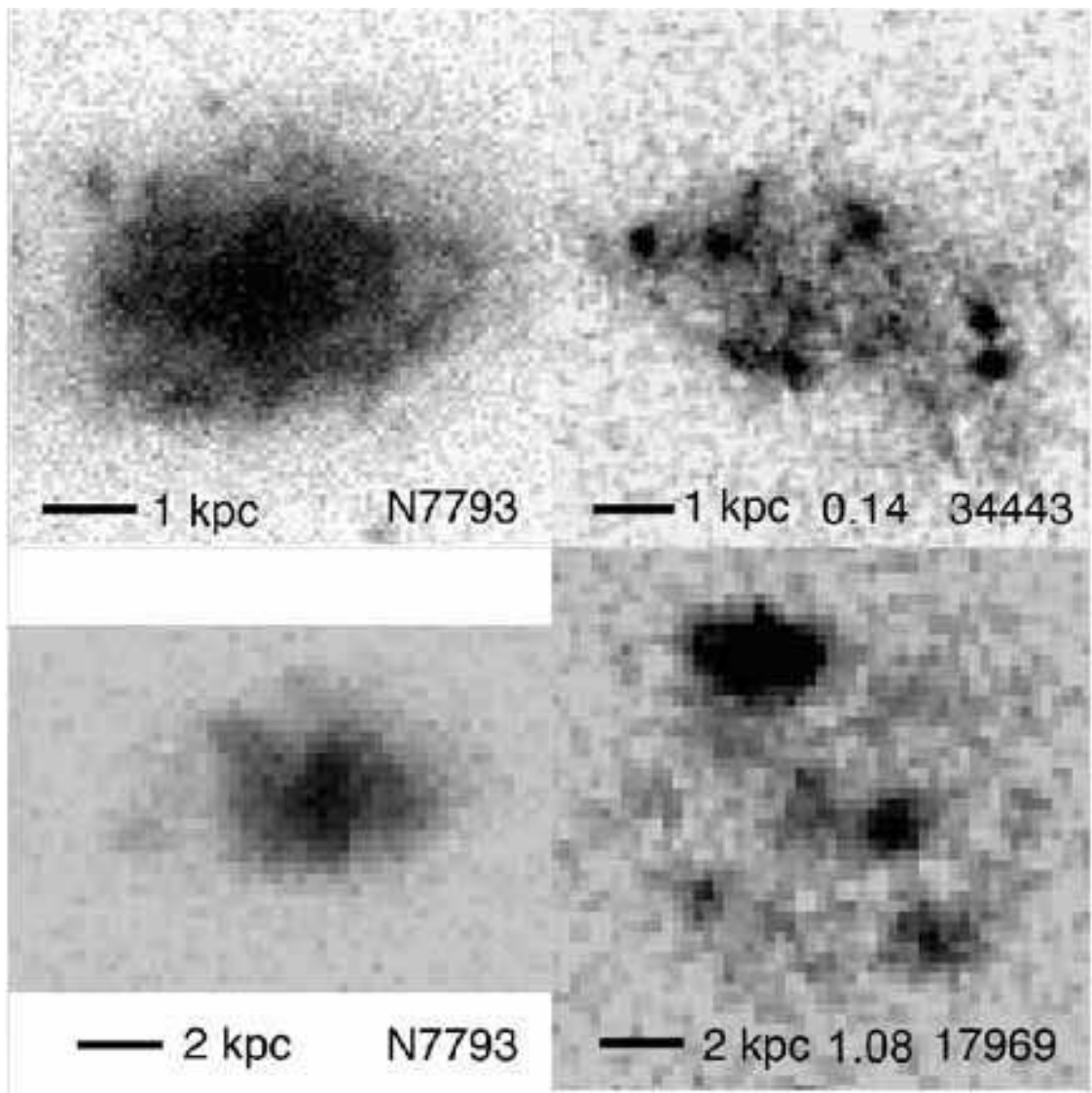} \caption{Blurred images of the flocculent galaxy
NGC 7793 are shown on the left using the same restframe wavelengths as
the GOODS images of two clump clusters, 34443 ($z=0.139$, top) and
17969 ($z=1.08$, bottom), shown on the right. The top left image of NGC
7793 is with a IIIa-J emulsion at 3950 \AA\ and has a spatial
resolution (FWHM) of 230 pc. The top right image of 34443 is in
$B_{435}$ and has the same rest wavelength and spatial resolution. The
bottom left image of NGC 7793 is in the NUV at 2267 \AA\ and has a
spatial resolution of 790 pc. The lower right image of 17969 is in
$B_{435}$ and has the same rest wavelength and spatial resolution. The
image of NGC 7793 was degraded to give the same pixel scale and
relative background noise as the corresponding clump cluster. Clump
cluster galaxies have higher clump-interclump contrasts than local
flocculent galaxies. [image degraded for arXiv preprint]
}\label{fig:fig19-N7793,GOODS-060409}\end{figure}

\clearpage
\begin{figure}\epsscale{1}
\plotone{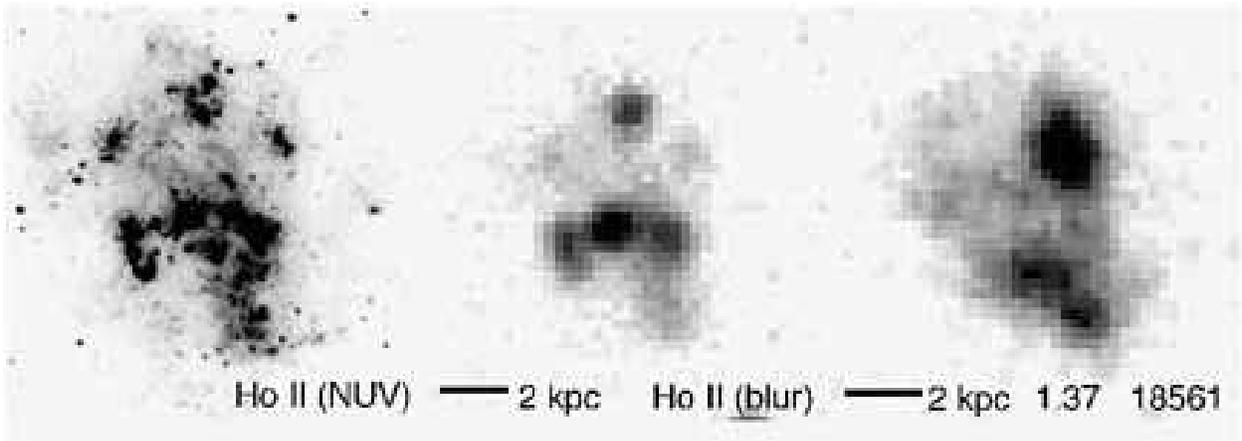}\caption{A NUV image of the local dwarf Irregular
Ho II is shown on the left and a blurred version is in the middle for
comparison to the GOODS clump cluster 18561, seen in the $V_{606}$ band
on the right. The spatial resolution (780 pc) and restframe wavelength
($\sim2400$ \AA) of the blurred image of Ho II and the ACS image of
18561 are the same. Local dwarf Irregular galaxies like Ho II resemble
high redshift clump cluster galaxies in their asymmetry and clumpy
structure. The local Irregulars are lower in mass by a factor of 10 to
100. [image degraded for arXiv
preprint]}\label{fig:HoII,18561-060409}\end{figure}

\end{document}